\documentclass{aa}

\usepackage{graphicx}
\usepackage{psfig}
\usepackage{txfonts}

\begin{document}

   \title{Efficiency of ETV diagrams as diagnostic tools for long-term period variations}

   \subtitle{II. Non-conservative mass transfer, and gravitational radiation}

   \author{N. Nanouris\inst{1,2,3},
           A. Kalimeris\inst{2},
           E. Antonopoulou\inst{1}
         \and
           H. Rovithis-Livaniou\inst{1}
           }

   \institute{Department of Astrophysics, Astronomy and Mechanics,
   Faculty of Physics, National and Kapodistrian University
   of Athens, Panepistimiopolis Zografou, 15 784 Athens, Greece\\
              \email{nanouris;eantonop;elivan@phys.uoa.gr}
         \and
             Department of Environment Technologists, School of Technological Applications,
             Technological and Educational Institute of
             Ionian Islands, Panagoula, 29 100 Zakynthos, Greece\\
             \email{taskal@teiion.gr}
         \and
             Institute for Astronomy, Astrophysics, Space Applications and Remote Sensing,
             National Observatory of Athens, I. Metaxa \& Vas. Pavlou St., Palaia Penteli,
             15 236 Athens, Greece
           }

   \offprints{N.\,Nanouris\\ \email{nanouris@phys.uoa.gr}}

   \date{Received.../ Accepted...}


  \abstract
   {The credibility of an eclipse timing variation (ETV) diagram
   analysis is investigated for various manifestations of the mass
   transfer and gravitational radiation processes in binary systems.
   The monotonicity of the period variations and the morphology of the
   respective ETV diagrams are thoroughly explored in both the direct
   impact and the accretion disk mode of mass transfer, accompanied
   by different types of mass and angular momentum losses (through a
   hot-spot emission from the gainer and via the L2/L3 points).}
   {Our primary objective concerns the traceability of each physical
   mechanism by means of an ETV diagram analysis. Also, possible
   critical mass ratio values are sought for those transfer modes
   that involve orbital angular momentum losses strong enough to dictate
   the secular period changes even when highly competitive mechanisms
   with the opposite direction act simultaneously.}
   {The $\dot{J}-\dot{P}$ relation that governs the orbital evolution
   of a binary system is set to provide the exact solution for the
   period and the function expected to represent the subsequent eclipse
   timing variations. The angular momentum transport is parameterized
   through appropriate empirical relations, which are inferred from
   semi-analytical ballistic models. Then, we numerically determine the
   minimum temporal range over which a particular mechanism is rendered
   measurable, as well as the critical mass ratio values that signify
   monotonicity inversion in the period modulations.}
   {Mass transfer rates comparable to or greater than
   $10^{-8}\,M_\mathrm{\odot}\,\mathrm{yr^{-1}}$ are measurable for
   typical noise levels of the ETV diagrams, regardless of whether
   the process is conservative. However, the presence of a transient
   disk around the more massive component defines a critical mass ratio
   ($q_{\mathrm{cr}} \approx 0.83$) above which the period turns out
   to decrease when still in the conservative regime, rendering the
   measurability of the anticipated variations a much more complicated
   task. The effects of gravitational radiation proved to be rather
   undetectable, except for systems with physical characteristics
   that only refer to cataclysmic variables.}
   {The monotonicity of the period variations and the curvature of the
   respective ETV diagrams depend strongly on the accretion mode and the
   degree of conservatism of the transfer process. Unlike the hot-spot
   effects, the Lagrangian points L2 and L3 support very efficient routes
   of strong angular momentum loss. It is further shown that escape of
   mass via the L3 point -- when the donor is the less massive component --
   safely provides critical mass ratios above which the period is expected
   to decrease, no matter how intense the process is.}

   \keywords{      binaries: close --
                   accretion, accretion disks --
                   gravitational waves --
                   methods: miscellaneous}

   \authorrunning{N. Nanouris et al.}
   \titlerunning{}
   \maketitle


\section{Introduction}

It is well known that the information taken directly from the
observed orbital evolution of eclipsing binary stars is valuable
for studying crucial physical mechanisms related to the stellar
structure and evolution. In this direction, the analysis of ETV
(eclipse timing variations) or, equivalently, of O--C (observed
minus calculated) diagrams plays an increasingly important role
as a tool with which the orbital period variations of eclipsing
pairs are disclosed (e.g., Hilditch \cite{H01}; Sterken \cite{S05}).
In the framework of further mathematical treatment, these variations
can be used to probe evolutionary processes and estimate the
parameters connected to the underlying physical mechanisms. In this
context, the current series of papers aims to further evaluate the
efficiency of ETV diagrams and to specify their limits as a
diagnostic tool for long-term observed period variations.

The incorporated mathematical approach has already been presented
in detail in an earlier paper (Nanouris et al. \cite{NKARL11},
hereafter NKARL11). In NKARL11, the mathematical procedure is
implemented for wind-driven mass loss and magnetic braking (mainly
to estimate the minimum temporal intervals required for detecting
their signal in ETV diagrams). The morphology and the general
properties of the ETV diagrams under the separate or joint influence
of these specific mechanisms were also examined mostly in late-type
detached binaries.

In this study we continue and further elaborate on the research by
focusing on the effects the following three physical mechanisms
have on the ETV diagram modulation: (\emph{a}) conservative mass
transfer (MT) through the Lagrangian L1 point, (\emph{b})
non-conservative MT accompanied by mass and angular momentum loss
(either as a hot-spot emission from the gainer or through the L2
and L3 point), and (\emph{c}) angular momentum loss (AML) due to
gravitational radiation. A non-conservative MT era will also be
referred as ``liberal", as first introduced by Warner (\cite{W78})
and widely adopted in many subsequent studies (see, e.g., van Rensbergen
et al. \cite{VRDGDLM08} and references therein). Finally, as case
studies of particular interest, we examine the effects that the
combined action of the adopted physical mechanisms has on the ETV
diagram structure, searching for possible critical mass ratio values
above or below which the orbital AML dictates the secular period
changes.

\section{Mathematical procedure}

In this paragraph we briefly outline the basic equations and the
general methodology and notation developed in NKARL11. In
particular, $M_{\mathrm{1}}$, $R_{\mathrm{1}}$ and $M_{\mathrm{2}}$,
$R_{\mathrm{2}}$ are the masses and the radii of the primary and the
secondary component, respectively. Their mass ratio
$q=M_{\mathrm{2}}/M_{\mathrm{1}}$ is imposed henceforth to take
values below (or equal to) unity only, i.e., $0<q\leq1$.

The $\dot{J}-\dot{P}$ relation of Kalimeris \& Rovithis-Livaniou
(\cite{KRL06}) is first adopted in order to associate the variations
in the state of the main physical parameters describing a binary
system with its orbital period. The relation generalizes the form of
Tout \& Hall (\cite{TH91}) by allowing mass losses (ML) from both
components, also taking the influence of their spin angular momenta
to the period variations into consideration (i.e., the spin-orbit
coupling).

Throughout our search, we assume for simplicity circular orbits
($e = 0$) with no developing departures from circularity at all
($\dot{e} = 0$), favoring mostly short- and intermediate-period
binaries as tidal evolution theory suggests (e.g., Koch \& Hrivnak
\cite{KH81}; Giuricin et al. \cite{GMM84}). Furthermore, wind-driven
ML (including magnetic braking for late-type components) are omitted
from the present analysis for convenience
(i.e., $\dot{M}_{\mathrm{w},i}=0$, $i = 1, 2$), aiming to constrict
our approach to the exchange processes that take place in a close pair,
reducing at the same time the free parameters of the physical problem.

According to these assumptions and simplifications, the generalized
$\dot{J}-\dot{P}$ relation of
Kalimeris \& Rovithis-Livaniou (\cite{KRL06}) is reduced to

\begin{equation}
\label{equation:1}
\frac{\dot{P}}{P}=-\frac{3(1-q)}{M_\mathrm{2}}\dot{M}_\mathrm{2}
-\frac{3q+2}{M_\mathrm{1}+M_\mathrm{2}}\dot{m}
-3\frac{\dot{J}_{\mathrm{1}}}{J_\mathrm{orb}}
-3\frac{\dot{J}_{\mathrm{2}}}{J_\mathrm{orb}}
+3\frac{\dot{J}}{J_\mathrm{orb}}.
\end{equation}

On the left-hand side (hereafter l.h.s.) of Eq.~(\ref{equation:1}),
$P = P(t)$ is the orbital period and $\dot{P}=\dot{P}(t)$ its rate
of change, while on the right-hand side (hereafter r.h.s.), the first
term refers to the MT through the L1 point, with
$\dot{M}_\mathrm{2}>0$ to declare the rate at which mass is accreted
by the gainer, supposing that the donor is the more massive star. ML
that accompanies non-conservative manifestations of the MT process is
also included with $\dot{m}<0$ to express the loss rate at which mass
is rejected by the secondary component and, subsequently, escapes from
the system.

When the donor is the less massive star, the approach may be easily
inverted in the opposite direction. Designating as
$\dot{M}_\mathrm{1}>0$ the rate at which mass is accreted by the gainer
and $\dot{m}<0$ the loss rate at which mass is rejected by the primary
component, Eq.~(\ref{equation:1}) is thus modified to the following
alternative form:

\begin{equation}
\label{equation:2}
\frac{\dot{P}}{P}=-\frac{3(q-1)}{M_\mathrm{2}}\dot{M}_\mathrm{1}
-\frac{3q^{-1}+2}{M_\mathrm{1}+M_\mathrm{2}}\dot{m}
-3\frac{\dot{J}_{\mathrm{1}}}{J_\mathrm{orb}}
-3\frac{\dot{J}_{\mathrm{2}}}{J_\mathrm{orb}}
+3\frac{\dot{J}}{J_\mathrm{orb}}.
\end{equation}

The third and the fourth terms in Eqs.~(\ref{equation:1})
and~(\ref{equation:2}) induct the spin angular momentum
contribution of each individual star by means of the
consequential -- advected -- angular momentum transport,
following the formalism of Gokhale et al. (\cite{GPF07}); i.e.,
$\dot{J}_\mathrm{1}=j_{\mathrm{1}}\dot{M}_{\mathrm{1}}$ and
$\dot{J}_\mathrm{2}=j_{\mathrm{2}}\dot{M}_{\mathrm{2}}$. More
particularly, in Eq.~(\ref{equation:1}), $j_{\mathrm{1}}$ and
$j_{\mathrm{2}}$ are the specific angular momenta of the matter
leaving the donor and the matter arriving at the accretor at rates
$\dot{M}_{\mathrm{1}}<0$ and $\dot{M}_{\mathrm{2}}>0$, respectively,
whereas the opposite framework with
$\dot{M}_{\mathrm{1}}>0$ and $\dot{M}_{\mathrm{2}}<0$ holds in
Eq.~(\ref{equation:2}).

Although the donor is not considered synchronized with the orbital
motion in Gokhale et al. (\cite{GPF07}), here it is assumed to be
tidally locked, following the approach of many earlier studies (e.g.,
Nelemans et al. \cite{NPZVY01}; Marsh et al. \cite{MNS04}). Even for
the gainer, tidal torques are disregarded from our approach owing to
the wide range of the synchronization timescales, dictated by the
friction mechanism and the intrinsic properties of the stellar outer
envelopes (Zahn \cite{Z77}; Zahn \cite{Z89}; Zahn \& Bouchet \cite{ZB89};
Tassoul \& Tassoul \cite{TT92a},b).

In Algols, for instance, tides seem incapable of spinning down the gainer
during the transfer process (Dervi\c{s}o\u{g}lu et al. \cite{DTI10};
Deschamps et al. \cite{DSDJ13}), considering that radiative damping is
the driving tidal dissipation mechanism for stars with a radiative envelope
(Zahn \cite{Z77}). By using the appropriate tidal torque and apsidal motion
constants of Zahn (\cite{Z75}, see Table 1 of his work) for the gainer of
\object{IO UMa} ($M_{\mathrm{1}}=2.1\,M_{\mathrm{\odot}}$, e.g.,
Soydugan et al. \cite{SSKL13}), for example, a long synchronization time of
$9.69\times10^{9}\,\mathrm{yr}$ arises. In stars possessing a convective
envelope, on the other hand, turbulent viscosity becomes the dominant
tidal-braking process, a much more effective mechanism than the radiative
dissipation (Zahn \cite{Z77}). Considering the low-mass accretor of the
system \object{RY Aqr} ($M_{\mathrm{1}}=1.27\,M_{\mathrm{\odot}}$, e.g.,
Helt \cite{H87}), which has fractional radius (with respect to the orbital)
similar to that of the gainer in \object{IO UMa} ($\sim 0.17$), a timescale that is
four orders of magnitude shorter, $5.23\times10^{5}\,\mathrm{yr}$, is
estimated. Tides may also play an important role in the spin rates of double
white dwarfs and cataclysmic variables. Although the viscosity of degenerate
matter suggests timescales on the order of $10^{15}\,\mathrm{yr}$, turbulent
dissipation and the excitation of the non-radial modes are able to shorten
the synchronization time in values even lower than 500\,yr (see Marsh et al.
\cite{MNS04} and references therein). As a result, neglecting tidal coupling
might affect our approach mostly when turbulent convection is the principal
friction process, expected in late-type stars and, possibly, in degenerate
components.

In both Eqs.~(\ref{equation:1}) and~(\ref{equation:2}), the fifth
r.h.s. term represents the angular momentum lost by the system (at
a rate $\dot{J}$), while $J_{\mathrm{orb}}$ accounts for the
orbital angular momentum:

\begin{equation}
\label{equation:3}
J_\mathrm{orb}=M_\mathrm{red}A_\mathrm{orb}^{2}\Omega_\mathrm{kep}.
\end{equation}

In Eq.~(\ref{equation:3}),
$M_\mathrm{red}=M_\mathrm{1}M_\mathrm{2}/(M_\mathrm{1}+M_\mathrm{2})$
is the reduced mass of the system,
$A_\mathrm{orb}=G^{1/3}(M_\mathrm{1}+M_\mathrm{2})^{1/3}P^{2/3}/(2\pi)^{2/3}$
is the orbital radius, $\Omega_{\mathrm{kep}}=2\pi/P$ is the
Keplerian angular velocity, and $G$ the gravitational constant.

The differential Eqs.~(\ref{equation:1}) and~(\ref{equation:2})
can be solved analytically for the orbital period function $P(t)$
when the action of each individual physical process (or any
combination) of our interest is given or even simulated. In
consequence, if $\epsilon$ denotes a real-valued -- continuous --
measure of the orbital cycle (which also includes integer values
when primary light minima are concerned, e.g., Gimenez \&
Garcia-Pelayo \cite{GGP83}; Borkovits et al. \cite{BEFTK03}), the
orbital period as a function of time, $P(t)$, is converted into
the orbital cycle domain, $P(\epsilon)$, through the $t =
t(\epsilon)$ transformation that emerges by solving for $t$ the
following equation:

\begin{equation}
\label{equation:4} \frac{\mathrm{d}t}{\mathrm{d}\epsilon}=P(t).
\end{equation}

In this way, the function $\Delta T(\epsilon)$ that practically
represents the synthetic (theoretically anticipated) ETV diagram
is eventually determined, supposing that the calculated times of
minima are supplied by a linear ephemeris with initial minimum
time $T_{\mathrm{0}}$ and reference period $P_{\mathrm{e}}$.
Indeed, as the $P(t)$ function is given by Eqs.~(\ref{equation:1})
or~(\ref{equation:2}) along with its cycle transformed counterpart
$P(\epsilon)$ (through Eq.~(\ref{equation:4})), then
$\Delta T(\epsilon)$ can be specified by solving the following
differential equation (e.g., NKARL11):

\begin{equation}
\label{equation:5} \frac{\mathrm{d}\Delta
T(\epsilon)}{\mathrm{d}\epsilon}=P(\epsilon)-P_{\mathrm{e}}.
\end{equation}

All involved ordinary first-order differential equations (i.e.,
Eqs.~(\ref{equation:1})/(\ref{equation:2}), (\ref{equation:4}), and
(\ref{equation:5})) are solved under the initial conditions
$(\epsilon, t, P(\epsilon), \Delta T(\epsilon)) = (0,
T_{\mathrm{0}}, P_{\mathrm{e}}, 0)$ that eventually provide the
$\Delta T(\epsilon)$ function. If, additionally, $\varepsilon$
denotes the noise level of an ETV diagram, the critical number of
epochs $\epsilon_{\mathrm{min}}$, needed for the detection of a
signal induced by a certain mechanism, is then made available by
solving the equation,

\begin{equation}
\label{equation:6} \Delta T(\epsilon)=\varepsilon,
\end{equation}
which, alternatively, provides the number of epochs such that the
signal-to-noise ratio equals one.

Similarly to NKARL11, the value of $\varepsilon=0.001\,\mathrm{d}$
is considered here as a typical noise level associated with
photoelectric times of minima, $\varepsilon=0.03\,\mathrm{d}$ is
adopted as a typical value associated with visual or patrol plates
times of minima, while $\varepsilon=0.01\,\mathrm{d}$ serves as a
pilot statistical noise level (i.e., for cases where the presence of
migrating spots cannot be excluded, see also Kalimeris et al.
\cite{KRLR02}).

\section{Examination of physical mechanisms}

The implementation of the preceding methodology follows in four
sections. In the first we examine the effects that MT through the
L1 point (including those induced by a hot-spot emission) has on
ETV diagrams, while the second section considers the effects of ML
through the outer Lagrangian points L2 and L3. In the third section,
we design non-conservative MT schemes that unveil the ETV diagram
curvature based on the mass ratio of a system for a certain degree
of conservation. Finally, in a fourth, the impact of AML via
gravitational radiation in the O--C modulation, is examined.

\subsection{Case I: Mass exchange through the L1 point}

A semi-detached configuration is attained as long as one of the
two components of a binary system reaches the limiting surface of
the Roche lobe. The MT process, also known as Roche lobe overflow
(RLOF), is then realized through a critical (i.e., gravitationally
neutral) point that coincides with the Lagrangian point L1 for a
fully synchronized system (e.g., Kuiper \cite{K41}; Kopal \cite{K56};
Plavec \cite{P58}; Kruszewski \cite{K63}).

The inflowing matter may either hit the gainer directly (hereafter
direct-impact mode), or it can orbit around the gainer, producing
internal particle collisions due to intersecting trajectories. A
gaseous ring is built in this way and eventually turns into an
accretion disk owing to the effects of the gaseous material bulk
viscosity (hereafter disk formation mode, e.g.,
Kruszewski \cite{K67}; K\v{r}\'{i}\v{z} \cite{K71}; Lubow \& Shu
\cite{LS75}; Lubow \& Shu \cite{LS76}; Papaloizou \& Pringle
\cite{PP77}; Lin \& Papaloizou \cite{LP79}; Shu \& Lubow \cite{SL81}).
The latter occurs when the detached component has smaller radius than
a critical value,
$R_{\mathrm{min}}=r_{\mathrm{min}}A_{\mathrm{orb}}$, with
$r_{\mathrm{min}}$ to be given as a function of the mass ratio of the
system:

\begin{description}
\item[$\bullet$]
\begin{equation}
\label{equation:7}
r_{\mathrm{min}}=0.04838+0.03601\log{q}+0.01229(\log{q})^{2},
\end{equation}
\end{description}
when the donor is the more massive component, and

\begin{description}
\item[$\bullet$]
\begin{equation}
\label{equation:8}
r_{\mathrm{min}}=0.04930-0.03387\log{q}+0.05915(\log{q})^{2},
\end{equation}
\end{description}
when the donor is the less massive component.

This approximation has emerged from the ballistic calculations
of Lubow \& Shu (\cite{LS75}, see Table 2 of their work) and is
valid for $0.07 \leq q \leq 1$ with accuracy close to 1\% when
the donor is the more massive component and better than 0.5\%
when the donor is the less massive component.

\subsubsection{Mass transfer in the absence of an accretion disk}

In the direct-impact MT mode, the flow stream coming from the donor
hits the gainer and dissipates without creating an accretion disk
when the radius of the gainer is much larger than the respective
critical value $R_{\mathrm{min}}$ (e.g., K\v{r}\'{i}\v{z} \cite{K71};
Lubow \& Shu \cite{LS75}). The stream strikes the gainer almost
perpendicular to its surface and generates enough turbulence to keep
the friction timescale short (Biermann and Hall \cite{BH73};
Kaitchuck et al. \cite{KHS85}). As a result, tidal interaction seems
efficient enough to keep both components in a strong coupling regime,
canceling the angular momentum flows within the binary (e.g., Hut and
Paczynski \cite{HP84}; Verbunt and Rappaport \cite{VR88}), i.e.
$\dot{J}_{\mathrm{2}}\approx-\dot{J}_{\mathrm{1}}$. The orbital
angular momentum then remains invariable, and the period varies as a
result of the mass redistribution between the two components.

In the fully conservative framework
$(\dot{m}=\dot{J}=0)$, the $\dot{J}-\dot{P}$ relation given by
Eqs.~(\ref{equation:1}) or~(\ref{equation:2}) is reduced to an
ordinary first-order differential equation of the form,

\begin{equation}
\label{equation:9} \dot{P}=w_{\mathrm{L1}}P,
\end{equation}
where, when the donor is the more massive star, $w_{\mathrm{L1}}$
is given by

\begin{description}
\item[$\bullet$]
\begin{equation}
\label{equation:10}
w_{\mathrm{L1}}=-3(1-q)\frac{\dot{M}_{\mathrm{2}}}{M_{\mathrm{2}}}<0,
\end{equation}
\end{description}
while, when the donor is the less massive star, $w_{\mathrm{L1}}$
is given by

\begin{description}
\item[$\bullet$]
\begin{equation}
\label{equation:11}
w_{\mathrm{L1}}=-3(q-1)\frac{\dot{M}_{\mathrm{1}}}{M_{\mathrm{2}}}>0.
\end{equation}
\end{description}

Similar to NKARL11, the coefficient of proportionality between the
rate at which period changes and the period itself, $w_{\mathrm{L1}}$,
is considered constant since $\dot{M}_{i}$ terms are small enough to
affect the absolute parameters of any of the components during the
time windows covered even by the longest available ETV diagrams.
Therefore, solution of Eq.~(\ref{equation:9}) becomes trivial and
under the initial condition
$(t, P(t)) = (T_{\mathrm{0}}, P_{\mathrm{e}})$ is written as

\begin{equation}
\label{equation:12}
P(t)=P_\mathrm{e}\mathrm{e}^{w_{\mathrm{L1}}(t-T_\mathrm{0})}.
\end{equation}

Then, integration of Eq.~(\ref{equation:4}) under the initial
condition $(\epsilon,t)=(0,T_\mathrm{0})$ yields

\begin{equation}
\label{equation:13}
t-T_\mathrm{0}=-\frac{1}{w_{\mathrm{L1}}}\ln(1-w_{\mathrm{L1}}P_\mathrm{e}\epsilon).
\end{equation}

As a result, through Eqs.~(\ref{equation:12})
and~(\ref{equation:13}), the orbital period $P(\epsilon)$ as a
function of the continuous time variable $\epsilon$ is

\begin{equation}
\label{equation:14}
P(\epsilon)=\frac{P_\mathrm{e}}{1-w_{\mathrm{L1}}P_\mathrm{e}\epsilon}.
\end{equation}
Subsequently, integration of Eq.~(\ref{equation:5}) under
the initial condition $(\epsilon, \Delta T(\epsilon)) = (0, 0)$
eventually gives the corresponding solution for the $\Delta
T(\epsilon)$ function:

\begin{equation}
\label{equation:15} \Delta
T(\epsilon)=-\frac{1}{w_{\mathrm{L1}}}\ln(1-w_{\mathrm{L1}}P_\mathrm{e}\epsilon)-P_\mathrm{e}\epsilon.
\end{equation}

Equation~(\ref{equation:15}) implies that the curvature of
the $\Delta T(\epsilon)$ function, and subsequently the
anticipated form of the respective ETV diagram, is determined
by the sign of the constant $w_{\mathrm{L1}}$. In particular,
when the donor is the more massive star ($w_{\mathrm{L1}}<0$,
see Eq.~(\ref{equation:10})), the orbital period decreases
(see Eq.~(\ref{equation:9})) so the $\Delta T(\epsilon)$
function is concave for any $\epsilon$ (see Eq.~(\ref{equation:11})).
Likewise, when the donor is the less massive star
($w_{\mathrm{L1}}>0$), the orbital period increases while the
$\Delta T(\epsilon)$ function is convex for any $\epsilon$.
Therefore, the foregoing analysis recovers the widely used
procedure for estimating MT rates under conservative conditions
in semi-detached and contact binaries (e.g., Kwee \cite{K58};
Huang \cite{H63}; Kruszewski \cite{K64b}).

\begin{figure*}
\centering
\begin{tabular}{cc}
\includegraphics[width=8.8cm]{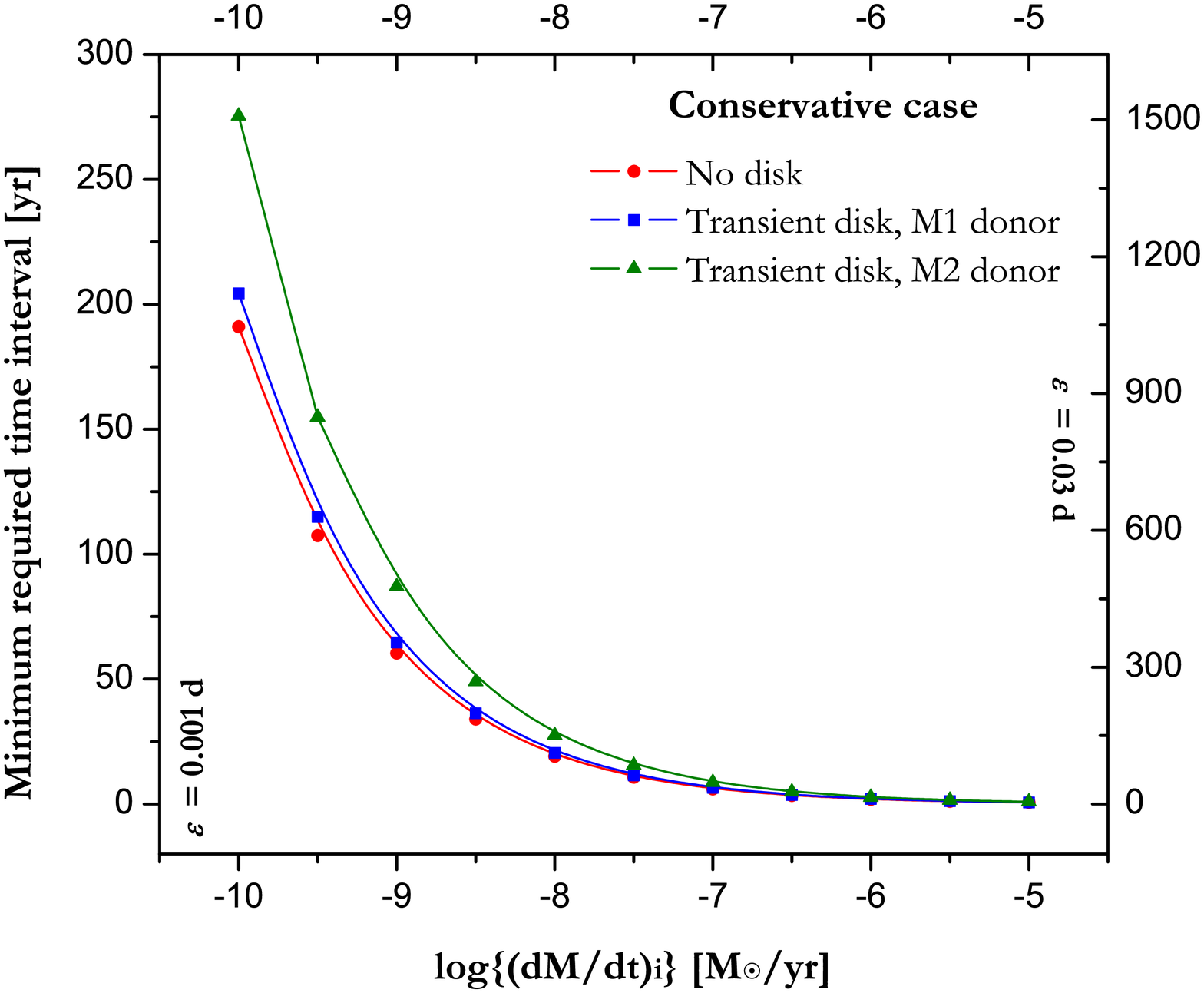}&\includegraphics[width=8.8cm]{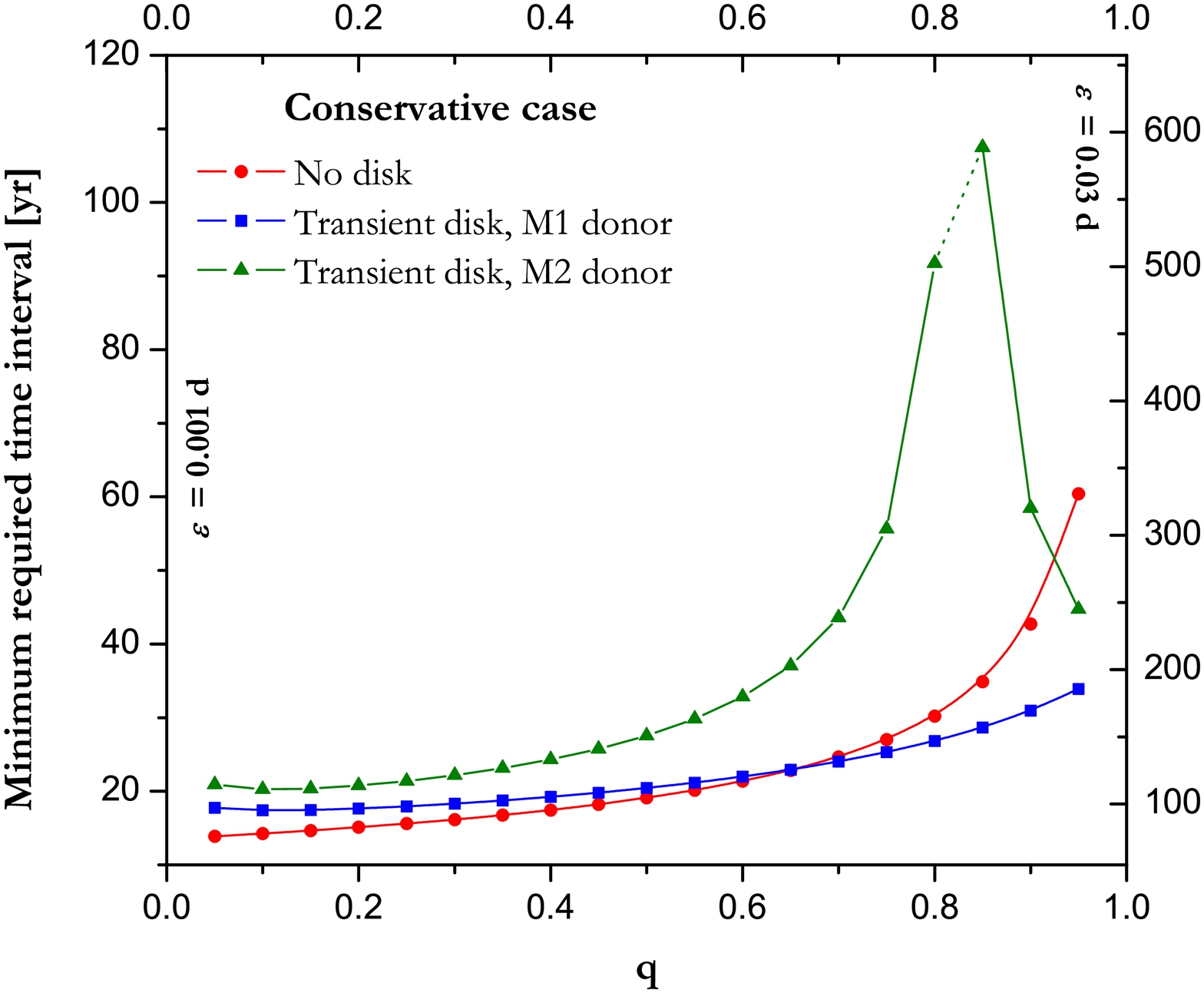}\\
(a)&(b)\\
\end{tabular}
\caption{Minimum time spans in a conservative era \textbf{a)} for
several mass transfer rates when the mass ratio is fixed at $q = 0.5$
and \textbf{b)} for various mass ratios when the mass transfer rate
is fixed at $10^{-8}\,M_\mathrm{\odot}\,\mathrm{yr}^{-1}$.
Computations were performed both in the absence (circles) and in the
presence of a transient disk around either the less massive (squares)
or the more massive component (up triangles), for noise levels equal
to 0.001\,d (left vertical axis) and 0.03\,d (right vertical axis).
The dashed line in the right panel covers the range within which the
critical value $q_{\mathrm{cr}} \approx 0.83$ lies. (This figure is
available in color in the electronic form.)
              }
         \label{figure:1}
   \end{figure*}

Given the $\Delta T(\epsilon)$ function, Eq.~(\ref{equation:6})
is solved by a suitable Newton-Raphson numerical scheme (e.g.,
Press et al. \cite{PTVF92}) in order to determine the minimum
epoch $\epsilon_{\mathrm{min}}$ and the subsequent temporal
intervals $t(\epsilon_{\mathrm{min}})-T_{\mathrm{0}}$ from
Eq.~(\ref{equation:13}). Our test case concerns a $q = 0.5$
semi-detached binary with a main-sequence secondary gainer of
fractional radius
$r_{\mathrm{2}} = R_{\mathrm{2}}/A_{\mathrm{orb}} = 0.17$
or a main-sequence primary gainer of fractional radius
$r_{\mathrm{1}} = R_{\mathrm{1}}/A_{\mathrm{orb}} = 0.25$,
keeping the secondary as massive as before.
Results for several MT rates with a pilot noise level of
$\varepsilon = 0.01\,\mathrm{d}$ (in Eq.~(\ref{equation:6}))
are given in Table~\ref{table:1}, while for
$\varepsilon = 0.001\,\mathrm{d}$ and $0.03\,\mathrm{d}$, the
expected intervals are depicted in Fig.~\ref{figure:1}a. Similar
computations were also made for various $q$ values at
$\dot{M}_{1}=10^{-8}\,M_\mathrm{\odot}\,\mathrm{yr}^{-1}$
(see Fig.~\ref{figure:1}b), always taking into consideration that
$r_{\mathrm{2}}$, $r_{\mathrm{1}} \gg r_{\mathrm{min}}$ should
be valid. The temporal intervals are expected to be the same
regardless of the component from which the mass begins to flow (given
that $|1-q|$ remains constant) and that they do not depend
on the reference period as long as Eq.~(\ref{equation:9}) holds
(see NKARL11).

\begin{table}
\caption{Minimum time spans at $\varepsilon = 0.01\,\mathrm{d}$ in
absence of an accretion disk (conservative case).}
\centering \label{table:1}
\begin{tabular}{c c |c c c c}
\hline \hline \multicolumn{2}{c |}{$q=0.5$}&
\multicolumn{4}{c}{$\dot{M}_{i}=10^{-8}\,M_\mathrm{\odot}\,\mathrm{yr^{-1}}$}\\
$\dot{M}_{i}$ & $t-T_\mathrm{0}$ & $~~~~q$ & $r_{\mathrm{2}}$ & $r_{\mathrm{1}}$ & $t-T_\mathrm{0}$\\
$[M_\mathrm{\odot}\,\mathrm{yr^{-1}}]$ & [yr] & & & & [yr]\\
\hline
$10^{-12}$ &  $6042  $ &   $~~~~0.9$ & 0.19  & 0.20 &  $135   $ \\
$10^{-11}$ &  $1911  $ &   $~~~~0.8$ & 0.18  & 0.21 &  $95.5  $ \\
$10^{-10}$ &  $604   $ &   $~~~~0.7$ & 0.18  & 0.22 &  $78.0  $ \\
$10^{-9} $ &  $191   $ &   $~~~~0.6$ & 0.17  & 0.24 &  $67.6  $ \\
$10^{-8} $ &  $60.4  $ &   $~~~~0.5$ & 0.17  & 0.25 &  $60.4  $ \\
$10^{-7} $ &  $19.1  $ &   $~~~~0.4$ & 0.16  & 0.28 &  $55.2  $ \\
$10^{-6} $ &  $6.0   $ &   $~~~~0.3$ & 0.15  & 0.31 &  $51.1  $ \\
$10^{-5} $ &  $1.9   $ &   $~~~~0.2$ & 0.13  & 0.36 &  $47.8  $ \\
$10^{-4} $ &  $0.6   $ &   $~~~~0.1$ & 0.11  & 0.45 &  $45.0  $ \\
\hline
\end{tabular}
\end{table}

Because most of the available times of minima span slightly more than
a century, a close inspection of Fig.~\ref{figure:1} shows that under
a low noise level of $\varepsilon = 0.001\,\mathrm{d}$, the signal of
the conservative MT process without disk formation can be detected at
rates $\dot{M}_{i}$ greater than
$10^{-10}\,M_\mathrm{\odot}\,\mathrm{yr}^{-1}$, and mass ratios even
close to unity. For higher noise levels, such as
$\varepsilon = 0.01\,\mathrm{d}$ or $0.03\,\mathrm{d}$, the signal is
rendered measurable only when
$\dot{M}_{i}>10^{-9}\,M_\mathrm{\odot}\,\mathrm{yr}^{-1}$ and $q < 0.9$
(see Table~\ref{table:1} and Fig.~\ref{figure:1}). The aforementioned
ascertainments are in fairly good agreement with MT rates derived so
far by traditional approaches for different binary types with at least
one component filling its Roche lobe, ranging from
$10^{-8}\,M_\mathrm{\odot}\,\mathrm{yr}^{-1}$ to
$10^{-5}\,M_\mathrm{\odot}\,\mathrm{yr}^{-1}$ (e.g., Kasz\'{a}s et
al. \cite{KVS98}; \v{S}imon \cite{S99}; Qian \& Liu \cite{QL00};
Qian \cite{Q02}; Qian \& Boonrucksar \cite{QB02}; Qian et al.
\cite{QYL07}; Manzoori \cite{M08}; Qian et al. \cite{QLFL08};
Zasche et al. \cite{ZLWN08}; Zhu et al. \cite{ZQLZM09};
Christopoulou et al. \cite{CPC11}).

\subsubsection{Mass transfer in the presence of a transient accretion disk}

Still in the direct-impact MT mode, the stream may be deflected
by the Coriolis force before striking the gainer when its radius
is marginally above the respective critical value $R_{\mathrm{min}}$
(e.g., K\v{r}\'{i}\v{z} \cite{K71}; Lubow \& Shu \cite{LS75}). The
stream hits the gainer tangentially to its surface, torquing the
star near its limb and producing a thin ring that is gradually
transformed into a quasi-stationary disk structure
(Kaitchuck et al. \cite{KHS85}; Peters \cite{P89};
Richards \& Albright \cite{RA99}). The disk will be considered as
transient (or low viscous) when the bulk viscosity is not efficient
enough to return the proffered angular momentum to the orbit (we
adopt the notation of Kaitchuck et al. \cite{KHS85}) or,
equivalently, when it experiences a weak coupling regime with the
donor. The advected angular momentum progressively accumulates in the
disk and spins up the gainer, forcing the orbital angular momentum to
decrease as a result of the conservative status (see, for instance,
Packet \cite{P81} for an excellent review on this topic).

In the present work, the specific angular momentum arriving at the
accretor is parameterized in terms of the circularization radius
(e.g., Biermann \& Hall \cite{BH73}; Flannery \cite{F75};
Warner \cite{W78}; Hut and Paczynski \cite{HP84};
Verbunt \& Rappaport \cite{VR88}), i.e. the equivalent radius of a
ring $R_{\mathrm{r}} = r_{\mathrm{r}}A_{\mathrm{orb}}$ in a circular
Keplerian orbit around the gainer that has the same specific angular
momentum as the transferred mass. Hence,
$j_{\mathrm{2}} = \sqrt{GM_{\mathrm{2}}R_{\mathrm{r}}}$ and
$j_{\mathrm{1}} = \sqrt{GM_{\mathrm{1}}R_{\mathrm{r}}}$ are
the specific angular momenta accumulated by the accretor as secondary
or primary, respectively, resulting in angular momentum gain at rate
$\dot{J}_{\mathrm{2}}/J_{\mathrm{orb}}=\sqrt{q(q+1)r_{\mathrm{r}}}\dot{M}_{\mathrm{2}}/M_{\mathrm{2}}$
and
$\dot{J}_{\mathrm{1}}/J_{\mathrm{orb}}=\sqrt{(q+1)r_{\mathrm{r}}}\dot{M}_{\mathrm{1}}/M_{\mathrm{2}}$
for each corresponding case.

Here, we provide an empirical relation for the fractional radius
of the ring $r_{\mathrm{r}}$ as a function of the mass ratio $q$,
inferred from the results of the semi-analytical approach of Lubow
\& Shu (\cite{LS75}) for sufficiently low initial flow stream
velocities at the L1 point (see Table 2 of their work):

\begin{description}
\item[$\bullet$]
\begin{equation}
\label{equation:16}
r_{\mathrm{r}}=0.08586+0.06214\log{q}+0.02022(\log{q})^{2},
\end{equation}
\end{description}
when the donor is the more massive component, and

\begin{description}
\item[$\bullet$]
\begin{equation}
\label{equation:17}
r_{\mathrm{r}}=0.08658-0.06925\log{q}+0.07368(\log{q})^{2},
\end{equation}
\end{description}
when the donor is the less massive component.

This approximation is valid for $0.07 \leq q \leq 1$, with the average
relative error of $r_{\mathrm{r}}$ being estimated close to 1\% when
the donor is the more massive component and less than 0.5\% when the
less massive. In the latter case, we also refer to the even more accurate
relation of Verbunt \& Rappaport (\cite{VR88}) when a wider mass ratio
range is needed (i.e., for $0.001 \leq q \leq 1$).

As far as the specific angular momentum leaving the donor is concerned,
if $R_{\mathrm{L}} = r_{\mathrm{L}}A_{\mathrm{orb}}$ is the (equivalent)
radius of its Roche lobe, for the purposes of the present study we
simply adopt
$j_{\mathrm{1}} \approx R_{\mathrm{L}}^{2}\Omega_{\mathrm{kep}}$ in
Eq.~(\ref{equation:1}) and
$j_{\mathrm{2}} \approx R_{\mathrm{L}}^{2}\Omega_{\mathrm{kep}}$ in
Eq.~(\ref{equation:2}), following the suggestion of Gokhale et al.
(\cite{GPF07}) and recalling that the donor here is synchronized with the
orbital velocity and fills its Roche lobe. As a consequence, its inertia
offers an extra store of angular momentum, leaving the donor at a rate
equal to
$\dot{J}_{\mathrm{1}}/J_{\mathrm{orb}}=-(q+1)r_{\mathrm{L}}^{2}\dot{M}_{\mathrm{2}}/M_{\mathrm{2}}$
and
$\dot{J}_{\mathrm{2}}/J_{\mathrm{orb}}=-(q+1)r_{\mathrm{L}}^{2}\dot{M}_{\mathrm{1}}/M_{\mathrm{2}}$
for each case, respectively, considering that
$\dot{M}_{\mathrm{1}}=-\dot{M}_{\mathrm{2}}$ in the conservative framework.
The fractional Roche lobe radius $r_{\mathrm{L}}$ is estimated via the
empirical law of Eggleton (\cite{E83}) with accuracy better than 1\%,
properly modified so that $q \leq 1$ (e.g., Hilditch \cite{H01}):

\begin{description}
\item[$\bullet$]
\begin{equation}
\label{equation:18}
r_{\mathrm{L}}=\frac{0.49q^{-2/3}}{0.6q^{-2/3}+\ln(1+q^{-1/3})},
\end{equation}
\end{description}
when the donor is the more massive component, and

\begin{description}
\item[$\bullet$]
\begin{equation}
\label{equation:19}
r_{\mathrm{L}}=\frac{0.49q^{2/3}}{0.6q^{2/3}+\ln(1+q^{1/3})},
\end{equation}
\end{description}
when the donor is the less massive component.

As a consequence, the $\dot{J}-\dot{P}$ relation, given by
Eqs.~(\ref{equation:1}) and~(\ref{equation:2}), turns into an
ordinary first-order differential equation identical to
Eq.~(\ref{equation:9}), $\dot{P}=w_{\mathrm{L1,r}}P$, which can be
solved for the orbital period $P(t)$, by taking into account that,
when the donor is the more massive star, the constant
$w_{\mathrm{L1,r}}$ is given by

\begin{description}
\item[$\bullet$]
\begin{equation}
\label{equation:20}
w_{\mathrm{L1,r}}=-3\left[1-q+\sqrt{q(q+1)r_{\mathrm{r}}}-(q+1)r_{\mathrm{L}}^{2}\right]\frac{\dot{M}_{\mathrm{2}}}{M_{\mathrm{2}}}<0,
\end{equation}
\end{description}
while, when the donor is the less massive star,
$w_{\mathrm{L1,r}}$ is given by

\begin{description}
\item[$\bullet$]
\begin{equation}
\label{equation:21}
w_{\mathrm{L1,r}}=-3\left[q-1+\sqrt{(q+1)r_{\mathrm{r}}}-(q+1)r_{\mathrm{L}}^{2}\right]\frac{\dot{M}_{\mathrm{1}}}{M_{\mathrm{2}}}.
\end{equation}
\end{description}

Since the presence of a transient disk drains orbital angular momentum,
when the donor is the more massive component, the orbital period still
decreases with respect to the no-disk case
($w_{\mathrm{L1,r}}<0$, see Eq.~(\ref{equation:20})). However, the sign
of $w_{\mathrm{L1,r}}$ is not steady when the donor is the less massive
star (see Eq.~(\ref{equation:21})), so the period varies according to
the degree the viscosity of the transient disk facilitates the orbital
AML. High viscosity values favor a strong gainer-disk-orbit coupling
regime that returns most of the transferred angular momentum back to the
orbit; i.e., $w_{\mathrm{L1,r}}>0$ and $\dot{P}>0$, while low viscosity
values favor a weak coupling regime that is inadequate for restraining
the orbital AML; i.e., $w_{\mathrm{L1,r}}<0$ and $\dot{P}<0$. The extra
deposit of angular momentum stemming from the spin contribution of the
donor proves to be too weak to inhibit the orbital shrinkage.

In the simplified framework of Eqs.~(\ref{equation:16}),
(\ref{equation:17}), (\ref{equation:18}), and (\ref{equation:19}), the
sign of $w_{\mathrm{L1,r}}$ is determined solely by the mass ratio of the
system $q$. In particular, the unique root at $q = 0.83$ of the term
$q-1+\sqrt{(q+1)r_{\mathrm{r}}}-(q+1)r_{\mathrm{L}}^{2}$ inside the
brackets in Eq.~(\ref{equation:21}), defines a critical value
$q_{\mathrm{cr}}$ of the mass ratio $q$ so that

\begin{description}
\item[$\bullet$] when $q<q_{\mathrm{cr}}=0.83$, then $\dot{P}>0$
and $\Delta T(\epsilon)$ is convex;
\end{description}

\begin{description}
\item[$\bullet$] when $q>q_{\mathrm{cr}}=0.83$, then $\dot{P}<0$
and $\Delta T(\epsilon)$ is concave.
\end{description}

The detection of a critical value $q_{\mathrm{cr}}$ for the mass ratio
offers strong support to concave modulated ETV diagrams in
high-mass-ratio, semi-detached binaries when the less massive component
is the donor, contradicting the predictions of the conservative MT
process in the absence of a disk.

Determination of the minimum time interval required in order
to be the signal of the examined process observable in an ETV
diagram initially concerns a $q = 0.5$ semi-detached binary with a
main-sequence gainer of fractional radius
$r_{\mathrm{2}} = R_{\mathrm{2}}/A_{\mathrm{orb}} = 0.07$ or
$r_{\mathrm{1}} = R_{\mathrm{1}}/A_{\mathrm{orb}} = 0.09$ as
the secondary or primary components, respectively. Were computed
several MT rates for both the pilot noise level
$\varepsilon = 0.01\,\mathrm{d}$ (see Tables~\ref{table:2} and
\ref{table:3}) and the more extreme cases of
$\varepsilon = 0.001\,\mathrm{d}$ and $0.03\,\mathrm{d}$
(see Fig.~\ref{figure:1}a). The same procedure was repeated for
various $q$ values at a steady MT rate of
$10^{-8}\,M_\mathrm{\odot}\,\mathrm{yr}^{-1}$
(see Fig.~\ref{figure:1}b) by evaluating that the gainer's
fractional radius always lies between $r_{\mathrm{min}}$ and
$r_{\mathrm{r}}$, quantities that are also presented in
Tables~\ref{table:2} and \ref{table:3}.

\begin{table}
\caption{Minimum time spans at $\varepsilon = 0.01\,\mathrm{d}$ in
the presence of a transient disk around the less massive component
(conservative case).} \centering \label{table:2}
\begin{tabular}{c c |c c c c c}
\hline \hline \multicolumn{2}{c |}{$q=0.5$}&
\multicolumn{5}{c}{$\dot{M}_{\mathrm{2}}=10^{-8}\,M_\mathrm{\odot}\,\mathrm{yr^{-1}}$}\\
$\dot{M}_{\mathrm{2}}$ & $t-T_\mathrm{0}$ & $~~~q$ & $r_{\mathrm{min}}$ & $r_{\mathrm{2}}$ & $r_{\mathrm{r}}$ & $t-T_\mathrm{0}$\\
$[M_\mathrm{\odot}\,\mathrm{yr^{-1}}]$ & [yr] & & & & & [yr]\\
\hline
$10^{-12}$ &   $6462  $ &  $~~~~0.9$ & 0.05 & 0.07 & 0.08 & $97.2$ \\
$10^{-11}$ &   $2044  $ &  $~~~~0.8$ & 0.05 & 0.07 & 0.08 & $84.8$ \\
$10^{-10}$ &   $646   $ &  $~~~~0.7$ & 0.04 & 0.07 & 0.08 & $76.0$ \\
$10^{-9} $ &   $204   $ &  $~~~~0.6$ & 0.04 & 0.07 & 0.07 & $69.5$ \\
$10^{-8} $ &   $64.6  $ &  $~~~~0.5$ & 0.04 & 0.07 & 0.07 & $64.6$ \\
$10^{-7} $ &   $20.4  $ &  $~~~~0.4$ & 0.04 & 0.06 & 0.06 & $60.8$ \\
$10^{-6} $ &   $6.5   $ &  $~~~~0.3$ & 0.03 & 0.06 & 0.06 & $57.9$ \\
$10^{-5} $ &   $2.0   $ &  $~~~~0.2$ & 0.03 & 0.05 & 0.05 & $55.8$ \\
$10^{-4} $ &   $0.6   $ &  $~~~~0.1$ & 0.02 & 0.04 & 0.04 & $55.1$ \\
\hline
\end{tabular}
\end{table}

\begin{table}
\caption{Minimum time spans at $\varepsilon = 0.01\,\mathrm{d}$ in
the presence of a transient disk around the more massive component
(conservative case).} \centering \label{table:3}
\begin{tabular}{c c |c c c c c}
\hline \hline \multicolumn{2}{c |}{$q=0.5$}&
\multicolumn{5}{c}{$\dot{M}_{\mathrm{1}}=10^{-8}\,M_\mathrm{\odot}\,\mathrm{yr^{-1}}$}\\
$\dot{M}_{\mathrm{1}}$ & $t-T_\mathrm{0}$ & $~~~q$ & $r_{\mathrm{min}}$ & $r_{\mathrm{1}}$ & $r_{\mathrm{r}}$ & $t-T_\mathrm{0}$\\
$[M_\mathrm{\odot}\,\mathrm{yr^{-1}}]$ & [yr] & & & & & [yr]\\
\hline
$10^{-12}$ &   $8709 $ &  $~~~~0.9$ & 0.05 & 0.07 & 0.09 & $185  $ \\
$10^{-11}$ &   $2754 $ &  $~~~~0.8$ & 0.05 & 0.07 & 0.09 & $290  $ \\
$10^{-10}$ &   $871  $ &  $~~~~0.7$ & 0.06 & 0.08 & 0.10 & $138  $ \\
$10^{-9} $ &   $275  $ &  $~~~~0.6$ & 0.06 & 0.08 & 0.11 & $104  $ \\
$10^{-8} $ &   $87.1 $ &  $~~~~0.5$ & 0.06 & 0.09 & 0.11 & $87.1 $ \\
$10^{-7} $ &   $27.5 $ &  $~~~~0.4$ & 0.07 & 0.09 & 0.13 & $76.9 $ \\
$10^{-6} $ &   $8.7  $ &  $~~~~0.3$ & 0.08 & 0.11 & 0.14 & $70.1 $ \\
$10^{-5} $ &   $2.8  $ &  $~~~~0.2$ & 0.10 & 0.12 & 0.17 & $65.7 $ \\
$10^{-4} $ &   $0.9  $ &  $~~~~0.1$ & 0.14 & 0.15 & 0.23 & $64.1 $ \\
\hline
\end{tabular}
\end{table}

Although the inferred intervals
$t(\epsilon_{\mathrm{min}})-T_{\mathrm{0}}$ are found to be slightly
different with respect to those in the absence of a disk when the
donor is the primary component, they are severely broadened in the
opposing framework, requiring MT rates enhanced by at least one
order of magnitude (i.e., greater than $10^{-9}$ for $\varepsilon$
= 0.001\,d and $10^{-8}\,M_{\mathrm{\odot}}\,\mathrm{yr}^{-1}$ for
$\varepsilon$ = 0.03\,d). Beside this, the existence of a critical
mass ratio $q_{\mathrm{cr}} \approx 0.83$, around which the period
is expected to remain approximately invariable, makes detecting such
a signal in an ETV diagram even harder. In practice, systems with mass
ratio between 0.8 and 0.9 are expected to display no visible O--C
variations with any of the adopted noise levels.

\subsubsection{Mass transfer in the presence of a permanent accretion disk}

In the disk formation MT mode, the stream misses completely the
gainer when its radius is below the respective critical value
$R_{\mathrm{min}}$ (e.g., K\v{r}\'{i}\v{z} \cite{K71}; Lubow \& Shu
\cite{LS75}). As a consequence, a rotating ring is formed that
spreads out and is progressively transmuted into a stationary disk
structure (e.g., Papaloizou \& Pringle \cite{PP77}; Lin \& Papaloizou
\cite{LP79}). The disk will be considered as permanent (or
high viscous) provided that the fluid's material bulk
viscosity is efficient enough to return the proffered angular
momentum back to the orbit (once more, we adopt the notation of
Kaitchuck et al. \cite{KHS85}) or, equivalently, when it experiences
a strong coupling regime with the donor (e.g., Hut and Paczynski
\cite{HP84}; Verbunt and Rappaport \cite{VR88}). However, material at
the inner edge of the disk still has angular momentum that acts to
spin up the accretor at the expense of the orbital angular momentum
(e.g., Packet \cite{P81}; Marsh et al. \cite{MNS04}; Dervi\c{s}o\u{g}lu
et al. \cite{DTI10}; Deschamps et al. \cite{DSDJ13}).

The description of the period evolution of such systems does not differ
at all from those systems that accommodate a transient disk, by simply replacing
$r_{\mathrm{r}}$ by $r_{\mathrm{2}} = R_{\mathrm{2}}/A_{\mathrm{orb}}$ or
$r_{\mathrm{1}} = R_{\mathrm{1}}/A_{\mathrm{orb}}$ when the gainer is the
less or the more massive star, respectively (Marsh et al. \cite{MNS04}).
The minimum temporal intervals $t(\epsilon_{\mathrm{min}})-T_{\mathrm{0}}$
derived in the presence of a transient disk should be considered mostly as an
upper limit (see Tables~\ref{table:2} and \ref{table:3} and Fig.~\ref{figure:1}),
while the results inferred in the absence of a disk should mostly be regarded as
a lower limit (see Table~\ref{table:1} and Fig.~\ref{figure:1}), reflecting the
case where $r_{\mathrm{2}} \ll 1$ or $r_{\mathrm{1}} \ll 1$ when the gainer
is the secondary or the primary, respectively.

\subsubsection{Mass and angular momentum loss through a hot-spot emission}

When a semi-detached binary undergoes a MT phase, a hot spot is expected to
be created in the area that the inflowing material impacts the mass-gaining
component (whether an accretion disk is present or not, e.g.,
Kruszewski \cite{K64a}; Smak \cite{S71}; Warner \& Nather \cite{WN71};
Lubow \& Shu \cite{LS75}; Lubow \& Shu \cite{LS76}). As long as the radiative
energy of such a hot spot, strengthened due to the limited accreted area (along
with the rotational kinetic energy), surmounts the gravitational binding energy,
the system is subject to a short non-conservative era, generally expected soon
after the onset of the RLOF phase. It is proved that this condition is fulfilled
provided that a critical MT rate has been reached
(van Rensbergen et al. \cite{VRDGDLM08}). As a result, a hot spot can be a key
factor in impelling a binary to a non-conservative phase.

Here, the AML process is parameterized in terms of the orbital specific angular
momentum by adopting the formalism of Rappaport et al. (\cite{RJW82}):

\begin{equation}
\label{equation:22} \dot{J}_{\mathrm{hs}} =
j_{\mathrm{hs}}\dot{m}_{\mathrm{hs}}\Omega_{\mathrm{kep}}A_{\mathrm{orb}}^{2},
\end{equation}
where $\dot{m}_{\mathrm{hs}}<0$ denotes the ML rate during the non-conservative
era and $j_{\mathrm{hs}}$ the -- dimensionless -- specific angular momentum taken
away from the orbit of the accreting star; i.e., $j_{\mathrm{hs}} = (1/1+q)^{2}$,
if this is the secondary, and $j_{\mathrm{hs}} = (q/1+q)^{2}$, if this is the
primary star (de Mink et al. \cite{DMPH07}; Rensbergen et al. \cite{VRDGMJDL11}).
The AML term in Eqs.~(\ref{equation:1}) and~(\ref{equation:2}) is then equal to
$\dot{J}_{\mathrm{hs}}/J_{\mathrm{orb}}=j_{\mathrm{hs}}(q+1)\dot{m}_{\mathrm{hs}}/M_{\mathrm{2}}$.

Attempting to isolate the ML and AML effects caused by the presence of a hot spot
on the orbital evolution, we proceed by disregarding the MT term
(i.e., $\dot{M}_{i}=0$), and the spin angular momentum of both components as well
(i.e., $\dot{J}_{\mathrm{2}}\approx-\dot{J}_{\mathrm{1}}$); in other words, we
explore the impact that a fully liberal era (in which a hot spot emits all the
incoming material) has on an ETV diagram morphology in the absence of a disk. The
differential equation that governs the orbital evolution of the system, as derived
from Eqs.~(\ref{equation:1}) and~(\ref{equation:2}), therefore has a similar form to
Eq.~(\ref{equation:9}), $\dot{P}=w_{\mathrm{L1,hs}}P$, with the $w_{\mathrm{L1,hs}}$
constant being given by the following relations:

\begin{description}
\item[$\bullet$]
\begin{equation}
\label{equation:23}
w_{\mathrm{L1,hs}}=\frac{-3q^{2}-2q+3}{q+1}\cdot\frac{\dot{m}_{\mathrm{hs}}}{M_{\mathrm{2}}},
\end{equation}
\end{description}
when the donor is the more massive component, and

\begin{description}
\item[$\bullet$]
\begin{equation}
\label{equation:24}
w_{\mathrm{L1,hs}}=\frac{3q^{2}-2q-3}{q+1}\cdot\frac{\dot{m}_{\mathrm{hs}}}{M_{\mathrm{2}}}>0,
\end{equation}
\end{description}
when the donor is the less massive component.

Equations~(\ref{equation:23}) and~(\ref{equation:24}) make it clear that the
prevalence of the driving mechanism (the ML process itself against the ML-induced
torque through which the system loses angular momentum) depends directly on the
mass ratio of the system. The sign investigation of $w_{\mathrm{L1,hs}}$ in
Eq.~(\ref{equation:23}) shows that for binaries whose the donor is the primary
star, the orbital period is expected to increase when $q = q_{\mathrm{cr}} > 0.72$,
otherwise the period decreases. In contrast, in binaries whose donor is the
secondary star, the orbital period is expected to increase regardless of the mass
ratio $q$ values. The above inference clearly shows that the AML effect
hardly compensates for the ML process, resulting in an increasing period in most
cases.

Determination of $t(\epsilon_{\mathrm{min}})-T_{\mathrm{0}}$ takes place following
the standard procedure for the test cases examined in the direct-impact mode
without a transient disk. The estimated temporal windows are quoted in
Tables~\ref{table:4} and \ref{table:5} for $\varepsilon = 0.01\,\mathrm{d}$, while
they are graphically displayed in Fig.~\ref{figure:2} for
$\varepsilon = 0.001\,\mathrm{d}$ and 0.03\,d. Peering at the tabulated values, one
notes that an O--C modulation is rendered observable for ML rates that exceed
$-10^{-9}\,M_{\mathrm{\odot}}\,\mathrm{yr}^{-1}$, barely favoring the case of the
secondary star as the donor. Moreover, a non-conservative era due to a hot-spot
emission from the secondary is expected to be measurable for almost any mass ratio
except for a very narrow range between 0.7 and 0.8 (where
$q_{\mathrm{cr}} \thickapprox 0.72$ lies).

\begin{table}
\caption{Minimum time spans at $\varepsilon = 0.01\,\mathrm{d}$
when a hot spot, located on the less massive component, emits all
the incoming material (fully liberal case).} \centering
\label{table:4}
\begin{tabular}{c c |c c c}
\hline \hline \multicolumn{2}{c |}{$q=0.5$}&
\multicolumn{3}{c}{$\dot{m}_{\mathrm{hs}}=-10^{-8}\,M_\mathrm{\odot}\,\mathrm{yr^{-1}}$}\\
$\dot{m}_{\mathrm{hs}}$ & $t-T_\mathrm{0}$ & $~~~~q$ & $r_{\mathrm{2}}$ & $t-T_\mathrm{0}$\\
$[M_\mathrm{\odot}\,\mathrm{yr^{-1}}]$ & [yr] & & & [yr]\\
\hline
$-10^{-12}$ &   $8106   $ &   $~~~~0.9$ & $0.19$ &  $92.0   $ \\
$-10^{-11}$ &   $2563   $ &   $~~~~0.8$ & $0.18$ &  $138    $ \\
$-10^{-10}$ &   $811    $ &   $~~~~0.7$ & $0.18$ &  $268    $ \\
$-10^{-9} $ &   $256    $ &   $~~~~0.6$ & $0.17$ &  $110    $ \\
$-10^{-8} $ &   $81.1   $ &   $~~~~0.5$ & $0.17$ &  $81.1   $ \\
$-10^{-7} $ &   $25.6   $ &   $~~~~0.4$ & $0.16$ &  $66.8   $ \\
$-10^{-6} $ &   $8.1    $ &   $~~~~0.3$ & $0.15$ &  $57.8   $ \\
$-10^{-5} $ &   $2.6    $ &   $~~~~0.2$ & $0.13$ &  $51.5   $ \\
$-10^{-4} $ &   $0.8    $ &   $~~~~0.1$ & $0.11$ &  $46.6   $ \\
\hline
\end{tabular}
\end{table}

\begin{table}
\caption{Minimum time spans at $\varepsilon = 0.01\,\mathrm{d}$
when a hot spot, located on the more massive component, emits all
the incoming material (fully liberal case).} \centering
\label{table:5}
\begin{tabular}{c c |c c c}
\hline \hline \multicolumn{2}{c |}{$q=0.5$}&
\multicolumn{3}{c}{$\dot{m}_{\mathrm{hs}}=-10^{-8}\,M_\mathrm{\odot}\,\mathrm{yr^{-1}}$}\\
$\dot{m}_{\mathrm{hs}}$ & $t-T_\mathrm{0}$ & $~~~~q$ & $r_{\mathrm{1}}$ & $t-T_\mathrm{0}$\\
$[M_\mathrm{\odot}\,\mathrm{yr^{-1}}]$ & [yr] & & & [yr]\\
\hline
$-10^{-12}$ &   $5027   $ &   $~~~~0.9$ & $0.20$ &  $66.3   $ \\
$-10^{-11}$ &   $1590   $ &   $~~~~0.8$ & $0.21$ &  $60.6   $ \\
$-10^{-10}$ &   $503    $ &   $~~~~0.7$ & $0.22$ &  $56.4   $ \\
$-10^{-9} $ &   $159    $ &   $~~~~0.6$ & $0.24$ &  $53.0   $ \\
$-10^{-8} $ &   $50.3   $ &   $~~~~0.5$ & $0.25$ &  $50.3   $ \\
$-10^{-7} $ &   $15.9   $ &   $~~~~0.4$ & $0.28$ &  $48.1   $ \\
$-10^{-6} $ &   $5.0    $ &   $~~~~0.3$ & $0.31$ &  $46.2   $ \\
$-10^{-5} $ &   $1.6    $ &   $~~~~0.2$ & $0.36$ &  $44.8   $ \\
$-10^{-4} $ &   $0.5    $ &   $~~~~0.1$ & $0.45$ &  $43.6   $ \\
\hline
\end{tabular}
\end{table}

By reducing the noise level at $\varepsilon$ = 0.001\,d, ML rates stronger than
$-10^{-10}\,M_{\mathrm{\odot}}\,\mathrm{yr}^{-1}$ are still detectable for a
century, while no significant changes occur by worsening the noise contamination
at $\varepsilon$ = 0.03\,d (Fig.~\ref{figure:2}a). However, by keeping the ML rate
fixed at $\dot{m}_{\mathrm{hs}}=-10^{-8}\,M_{\mathrm{\odot}}\,\mathrm{yr}^{-1}$,
the O--C variations are visible for $\varepsilon$ = 0.001\,d regardless of the mass
ratio value, whereas a noise level increase at $\varepsilon$ = 0.03\,d limits the
traceability of the examined mechanism only for very small mass ratios when the
donor is expected to be the primary star (Fig.~\ref{figure:2}b).

\subsection{Case II: Mass and angular momentum loss through the L2/L3 points}

In the course of a violent MT process in a semi-detached binary (e.g., in
super-Eddington accretion flows), mass and angular momentum may be lost from the
system either through the L2 or the L3 point when the donor is the more or the less
massive component, respectively. These are the paths that matter can escape most
easily from the gravitational field of the binary since it overcomes the lowest
potential barrier, as a result requiring the lowest energy of all. The formation
process and the observed features of a circumbinary envelope due to outflows from
these two points have been confirmed or/and explored widely in literature
(Kuiper \cite{K41}; Shu et al. \cite{SAL79}; Soberman et al. \cite{SPV97};
Spruit \& Taam \cite{ST01}; Taam \& Spruit \cite{TS01}; Chen et al. \cite{CLQ06};
D'Souza et al. \cite{DMTF06}; Motl et al. \cite{MFTD07}; Sytov et al. \cite{SKBKB07};
Sytov et al. \cite{SBKB09a},b; Zhilkin \& Bisikalo \cite{ZB09}; Bisikalo \cite{B10};
Mennickent et al. \cite{MKGO10}; Mennickent et al. \cite{MKD12}).

Here we adopt the AML parametrization proposed by Shu et al. (\cite{SAL79}), which
expresses the angular momentum taken away from the system via the L2 and L3 points
in terms of the orbital specific angular momentum, similarly to the hot spot case.
If $\dot{m}_{\mathrm{L2}} < 0$, $\dot{m}_{\mathrm{L3}} < 0$ defines the ML rate of
the escaping matter from the L2 and L3 points with -- dimensionless -- specific
angular momenta $j_{\mathrm{L2}}$, $j_{\mathrm{L3}}$, then $\dot{J}_{\mathrm{L2}}$
and $\dot{J}_{\mathrm{L3}}$ are estimated through Eq.~(\ref{equation:22}) by
properly substituting $\dot{m}_{\mathrm{hs}}$ and $j_{\mathrm{hs}}$ for each
corresponding case.

Intending to provide an efficient empirical relation for $j_{\mathrm{L2}}$ as a
function of the mass ratio $q$, we employ the values of the semi-analytical
ballistic model of  Shu et al. (\cite{SAL79}), as tabulated in Table 2 of their
study. The values account for the terminal specific angular momentum (at very large
distances from the system where the flow finally obtains its terminal kinematical
characteristics), also taking the oblique impact angle the outflowing stream
follows into account. In the absence of analytic calculations for $j_{\mathrm{L3}}$,
we assume that matter leaves the system at the L3 point in a direction vertical to
the line connecting the components (thus providing an upper AML limit); i.e.,
$j_{\mathrm{L3}} = x_{\mathrm{L3}}^2$, with $x_{\mathrm{L3}}$ denoting the relative
position of L3 in the Roche geometry coordinates system:

\begin{description}
\item[$\bullet$]
\begin{equation}
\label{equation:25}
j_{\mathrm{L2}}=1.70166-0.49695\log{q}-0.41981(\log{q})^{2},
\end{equation}
\end{description}
when ML and AML occur through L2, and

\begin{description}
\item[$\bullet$]
\begin{equation}
\label{equation:26}
j_{\mathrm{L3}}=1.43988+0.55460\log{q}+0.19061(\log{q})^{2},
\end{equation}
\end{description}
when ML and AML occur through L3.

This approximation is valid for $0.05 \leq q \leq 1$ with accuracy better than
0.5\% for both $j_{\mathrm{L2}}$ and $j_{\mathrm{L3}}$; however, $j_{\mathrm{L3}}$
represents the initial specific angular momentum (i.e., at L3), and a more reliable
error close to 15\% should be considered (as estimated by comparing the values of
initial and terminal specific angular momentum for $j_{\mathrm{L2}}$, which are
listed in Table 2 of Shu et al. \cite{SAL79}).

\begin{figure*}
\centering
\begin{tabular}{cc}
\includegraphics[width=8.8cm]{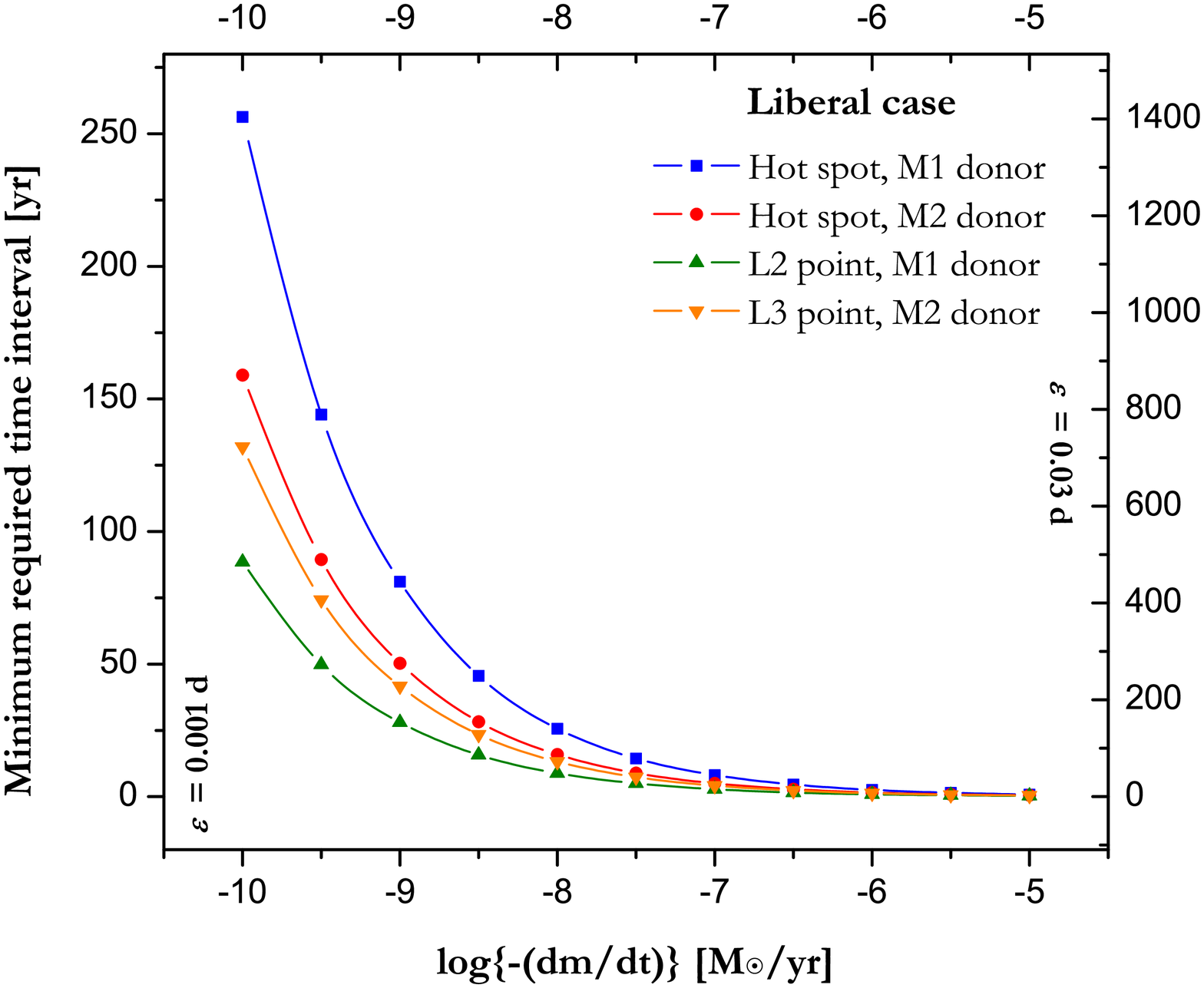}&\includegraphics[width=8.8cm]{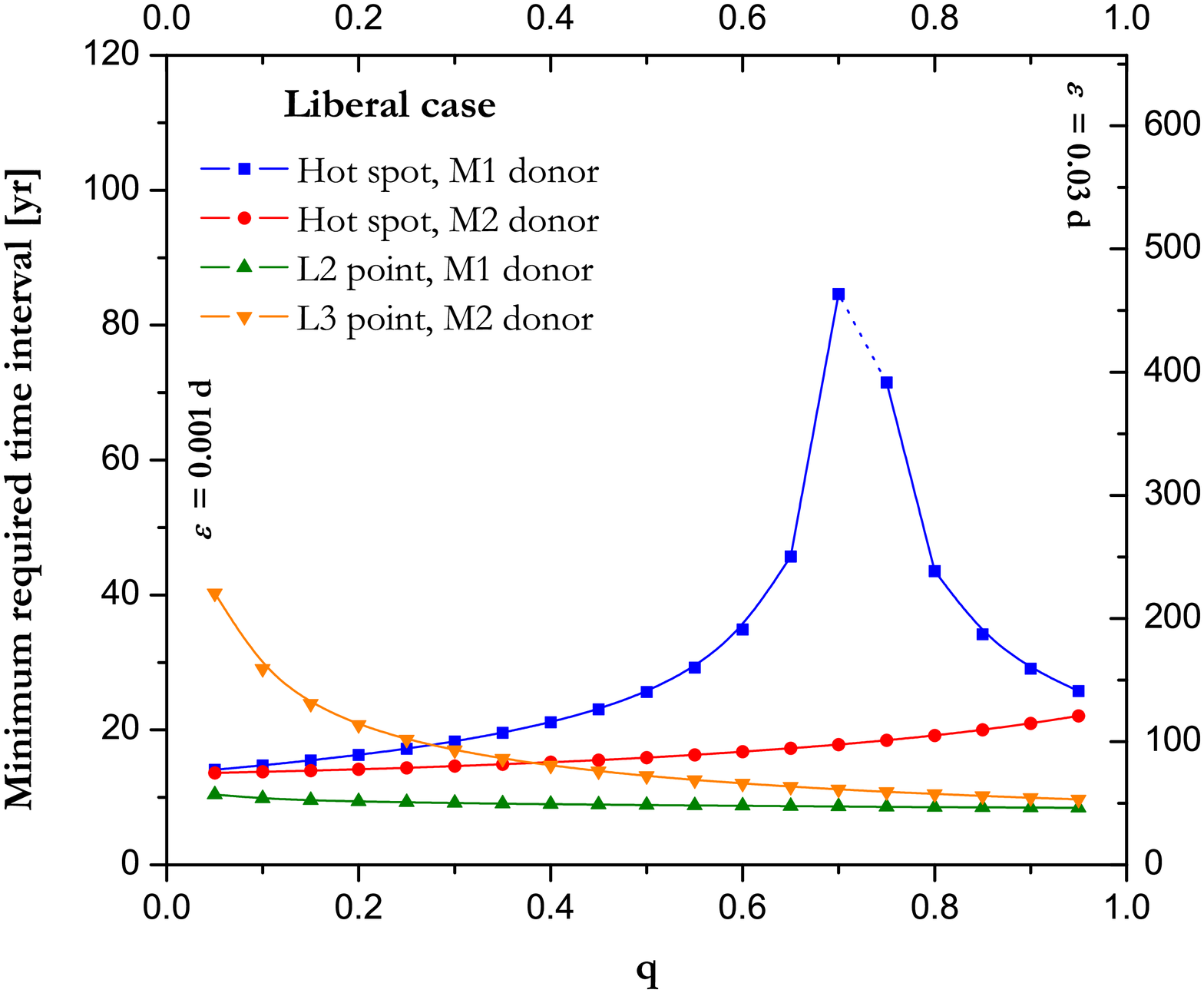}\\
(a)&(b)\\
\end{tabular}
\caption{Minimum time spans in a fully liberal era \textbf{a)} for
several mass loss rates when the mass ratio is fixed at $q = 0.5$
and \textbf{b)} for various mass ratios when the mass loss rate is
fixed at $-10^{-8}\,M_\mathrm{\odot}\,\mathrm{yr}^{-1}$.
Computations were performed considering that the transferred material
is lost from the system through a hot-spot emission, located either
on the less massive (squares) or the more massive component (circles),
and via the L2 (up triangles) or the L3 point (down triangles), for
noise levels equal to 0.001\,d (left vertical axis) and 0.03\,d (right
vertical axis). The dashed line in the right panel covers the range
within which the critical value $q_{\mathrm{cr}} \approx 0.72$ lies.
(This figure is available in color in the electronic form.)
              }
         \label{figure:2}
   \end{figure*}

Even after taking all assumptions mentioned in the hot spot case into consideration
and specifying that the AML terms $\dot{J}_{\mathrm{L2}}/J_{\mathrm{orb}}$ and
$\dot{J}_{\mathrm{L3}}/J_{\mathrm{orb}}$ in Eqs.~(\ref{equation:1})
and~(\ref{equation:2}) are equal to
$j_{\mathrm{L2}}(q+1)\dot{m}_{\mathrm{L2}}/M_{\mathrm{2}}$ and
$j_{\mathrm{L3}}(q+1)\dot{m}_{\mathrm{L3}}/M_{\mathrm{2}}$, respectively, the orbital
evolution of the system is then still described by Eq.~(\ref{equation:9}) by setting
the $w_{\mathrm{L2}}$ constant in place of $w_{\mathrm{L1}}$ as

\begin{description}
\item[$\bullet$]
\begin{equation}
\label{equation:27}
w_{\mathrm{L2}}=\frac{3j_{\mathrm{L2}}(q+1)^{2}-3q^{2}-2q}{q+1}\cdot\frac{\dot{m}_{\mathrm{L2}}}{M_{\mathrm{2}}}<0,
\end{equation}
\end{description}
when ML and AML occur through L2, and

\begin{description}
\item[$\bullet$]
\begin{equation}
\label{equation:28}
w_{\mathrm{L3}}=\frac{3j_{\mathrm{L3}}(q+1)^{2}-2q-3}{q+1}\cdot\frac{\dot{m}_{\mathrm{L3}}}{M_{\mathrm{2}}}<0,
\end{equation}
\end{description}
when ML and AML occur through L3.

It is noteworthy that the sign of the numerator appeared on the r.h.s. of
Eqs.~(\ref{equation:27}) and~(\ref{equation:28}) is positive for any mass ratio $q$
(since $j_{\mathrm{L2}}$, $j_{\mathrm{L3}} > 1$). Unlike the hot spot case,
specific angular momentum is carried away from the system at a rather distant point
(with respect to the center of mass), leading to markedly different AML rates and
warranting period reduction ($j_{\mathrm{hs}}<1$). As a result, the outer Lagrangian
points L2 and L3 turn out to be routes through which the excess of matter, surviving
the MT process, travels out of the system with prominent AML rates, ensuring orbital
shrinkage.

To examine the measurability of the ML mechanism through the L2 and the L3 points,
the standard methodology followed thus far is applied by retaining the binary
parameters used in the absence of a disk case. Accurate values regarding the minimum
intervals $t(\epsilon_{\mathrm{min}})-T_{\mathrm{0}}$ are given in Tables~\ref{table:6}
and \ref{table:7} for $\varepsilon$ = 0.01\,d, while they are graphically demonstrated
in Fig.~\ref{figure:2} for the two more noise levels of $\varepsilon$ = 0.001\,d and
0.03\,d.

\begin{table}
\caption{Minimum time spans at $\varepsilon = 0.01\,\mathrm{d}$
when material, ejected from the more massive component, is carried
away from the system through the L2 point (fully liberal case).}
\centering \label{table:6}
\begin{tabular}{c c |c c c}
\hline \hline \multicolumn{2}{c |}{$q=0.5$}&
\multicolumn{3}{c}{$\dot{m}_{\mathrm{L2}}=-10^{-8}\,M_\mathrm{\odot}\,\mathrm{yr^{-1}}$}\\
$\dot{m}_{\mathrm{L2}}$ & $t-T_\mathrm{0}$ & $~~~~q$ & $r_{\mathrm{2}}$ & $t-T_\mathrm{0}$\\
$[M_\mathrm{\odot}\,\mathrm{yr^{-1}}]$ & [yr] & & & [yr]\\
\hline
$-10^{-12}$ &   $2798   $ &   $~~~~0.9$ & $0.19$ &  $26.8   $ \\
$-10^{-11}$ &   $885    $ &   $~~~~0.8$ & $0.18$ &  $27.1   $ \\
$-10^{-10}$ &   $280    $ &   $~~~~0.7$ & $0.18$ &  $27.3   $ \\
$-10^{-9} $ &   $88.5   $ &   $~~~~0.6$ & $0.17$ &  $27.6   $ \\
$-10^{-8} $ &   $28.0   $ &   $~~~~0.5$ & $0.17$ &  $28.0   $ \\
$-10^{-7} $ &   $8.8    $ &   $~~~~0.4$ & $0.16$ &  $28.4   $ \\
$-10^{-6} $ &   $2.8    $ &   $~~~~0.3$ & $0.15$ &  $28.9   $ \\
$-10^{-5} $ &   $0.9    $ &   $~~~~0.2$ & $0.13$ &  $29.7   $ \\
$-10^{-4} $ &   $0.3    $ &   $~~~~0.1$ & $0.11$ &  $31.1   $ \\
\hline
\end{tabular}
\end{table}

Careful inspection of the results reveals that the O--C differences are detectable
as long as the ML rate from L2 exceeds
$-10^{-11}\,M_{\mathrm{\odot}}\,\mathrm{yr}^{-1}$ for $\varepsilon$ = 0.001\,d or $-10^{-10}\,M_{\mathrm{\odot}}\,\mathrm{yr}^{-1}$ when the noise level takes higher
values (i.e., $\varepsilon$ = 0.01 or 0.03\,d). However, the O--C residuals produced
by ML via the L3 point need a modestly longer baseline of observations compared to
the corresponding process via the L2 point. The mass ratio $q$ proves to be a factor
that hardly influences the temporal range $t(\epsilon_{\mathrm{min}})-T_{\mathrm{0}}$,
lying in a few decades no matter what the adopted noise level is.

\begin{table}
\caption{Minimum time spans at $\varepsilon = 0.01\,\mathrm{d}$
when material, ejected from the less massive component, is carried
away from the system through the L3 point (fully liberal case).}
\centering \label{table:7}
\begin{tabular}{c c |c c c}
\hline \hline \multicolumn{2}{c |}{$q=0.5$}&
\multicolumn{3}{c}{$\dot{m}_{\mathrm{L3}}=-10^{-8}\,M_\mathrm{\odot}\,\mathrm{yr^{-1}}$}\\
$\dot{m}_{\mathrm{L3}}$ & $t-T_\mathrm{0}$ & $~~~~q$ & $r_{\mathrm{1}}$ & $t-T_\mathrm{0}$\\
$[M_\mathrm{\odot}\,\mathrm{yr^{-1}}]$ & [yr] & & & [yr]\\
\hline
$-10^{-12}$ &   $4173   $ &   $~~~~0.9$ & $0.20$ &  $31.5   $ \\
$-10^{-11}$ &   $1320   $ &   $~~~~0.8$ & $0.21$ &  $33.2   $ \\
$-10^{-10}$ &   $417    $ &   $~~~~0.7$ & $0.22$ &  $35.4   $ \\
$-10^{-9} $ &   $132    $ &   $~~~~0.6$ & $0.24$ &  $38.2   $ \\
$-10^{-8} $ &   $41.7   $ &   $~~~~0.5$ & $0.25$ &  $41.7   $ \\
$-10^{-7} $ &   $13.2   $ &   $~~~~0.4$ & $0.28$ &  $46.6   $ \\
$-10^{-6} $ &   $4.2    $ &   $~~~~0.3$ & $0.31$ &  $53.8   $ \\
$-10^{-5} $ &   $1.3    $ &   $~~~~0.2$ & $0.36$ &  $65.7   $ \\
$-10^{-4} $ &   $0.4    $ &   $~~~~0.1$ & $0.45$ &  $92.0   $ \\
\hline
\end{tabular}
\end{table}

\subsection{Case III: Non-conservative mass transfer}

We examine now what the combined effects of MT, ML, and AML processes might be in the
modulation of an ETV diagram by distinguishing the case where the donor is the primary or
the secondary component. Furthermore, we discern two subcases of each non-conservative MT
case, addressing the general regime where a transient disk surrounds the gainer; the first
concerns ML and AML realized by a hot-spot emission, while the second regards ML and AML
from the L2/L3 point (non-captured mass from the gainer).

\subsubsection{Critical mass ratios when the donor is the more massive star}

If $\beta \in [0,1]$ is the fraction of the MT rate gained by the accretor, then $\beta = 1$
represents the conservative case (where all the transferred mass is captured from the gainer),
while $\beta = 0$ represents the fully liberal case (where all the transferred mass is rejected
from the gainer).

As far as the donor is the more massive component and provided that
$\dot{M}_{\mathrm{1}}=-\dot{M}_{\mathrm{2}}+\dot{m}$ is valid, the MT rate captured by the
secondary, $\dot{M}_{\mathrm{2}}$, and the remaining fraction that evades the system, $\dot{m}$,
can be written as

\begin{equation}
\label{equation:29}
\dot{M}_{\mathrm{2}}=-\beta\dot{M}_{\mathrm{1}},
\end{equation}
and
\begin{equation}
\label{equation:30} \dot{m}=(1-\beta)\dot{M}_{\mathrm{1}}.
\end{equation}

In the first subcase (hot-spot-driven ML and AML from the secondary), we combine
Eqs.~(\ref{equation:29}) and~(\ref{equation:30}) to express $\dot{M}_{\mathrm{1}}$ and
$\dot{m}_{\mathrm{hs}}$ as functions of the mass accretion rate $\dot{M}_{\mathrm{2}}$. Then,
Eq.~(\ref{equation:1}) reduces into the following form:

\begin{eqnarray}
\label{equation:31}
\frac{\dot{P}}{P}=[-3(1-q)&-&3\sqrt{q(q+1)r_{\mathrm{r}}}+\frac{3}{\beta}\cdot(q+1)r_{\mathrm{L}}^{2}\nonumber
\\&+&\frac{1-\beta}{\beta}\cdot\frac{3q^{2}+2q-3}{q+1}]\frac{\dot{M}_{\mathrm{2}}}{M_{\mathrm{2}}}.
\end{eqnarray}

Proceeding to the second subcase (ML and AML from L2), the rates $\dot{M}_{\mathrm{1}}$ and
$\dot{m}_{\mathrm{L2}}$ are expressed as a function of $\dot{M}_{\mathrm{2}}$ as well, and finally
Eq.~(\ref{equation:1}) takes the following form:

\begin{eqnarray}
\label{equation:32}
\frac{\dot{P}}{P}=[-3(1-q)&-&3\sqrt{q(q+1)r_{\mathrm{r}}}+\frac{3}{\beta}\cdot(q+1)r_{\mathrm{L}}^{2}\nonumber
\\&-&\frac{1-\beta}{\beta}\cdot\frac{3j_{\mathrm{L2}}(q+1)^{2}-3q^{2}-2q}{q+1}]\frac{\dot{M}_{\mathrm{2}}}{M_{\mathrm{2}}}.
\end{eqnarray}

It is worth noting that on the r.h.s. of Eqs.~(\ref{equation:31}) and~(\ref{equation:32}),
the expression within the brackets is a function of the mass ratio $q$ alone. As a
consequence, in the framework of our hypothesis set (and as long as $\beta$ is treated as
a free parameter), the mass ratio $q$ becomes a crucial factor that modulates the variations
(actually the monotonicity) of the period function $P(t)$, as well as the geometrical form
(actually the curvature) of the respective ETV diagram. The critical mass ratio values
$q_{\mathrm{cr}}$ can therefore arise after determining the roots of this expression, for any
$\beta$, considering the direct MT mode either in the presence of a transient disk orbiting
the secondary or in the absence of a disk. This can be achieved by calculating
$r_{\mathrm{r}}$ from Eq.~(\ref{equation:16}) and $r_{\mathrm{L}}$ from Eq.~(\ref{equation:18})
in the former case and by setting $r_{\mathrm{r}}=r_{\mathrm{L}}=0$ in the latter. Results
for $q_{\mathrm{cr}}$ are then presented in Table~\ref{table:8}. Binary systems with mass ratio
$q$ greater than the listed $q_{\mathrm{cr}}$ values will reveal an increasing period (and a
convex ETV diagram), as long as the driving processes are those examined in the current section.
New $q_{\mathrm{cr}}$ values can emerge by substituting $r_{\mathrm{2}}$ for $r_{\mathrm{r}}$
in the permanent disk formation mode.

\begin{table}
\caption{Critical mass ratios $q_{\mathrm{cr}}$ as a function of
the degree of liberalism $\beta$ when the donor is the more
massive component (see text for details).} \centering
\label{table:8}
\begin{tabular}{c c c | c c}
\hline \hline \multicolumn{3}{c |}{ML through a hot spot}&
\multicolumn{2}{c}{ML through the L2 point}\\
$\beta$ & $q_{\mathrm{cr}}$ & $q_{\mathrm{cr}}$ & $q_{\mathrm{cr}}$ & $q_{\mathrm{cr}}$\\
$$ & ($r_{\mathrm{r}},r_{\mathrm{L}}=0$) & ($r_{\mathrm{r}},r_{\mathrm{L}}\neq0$) & ($r_{\mathrm{r}},r_{\mathrm{L}}=0$) & ($r_{\mathrm{r}},r_{\mathrm{L}}\neq0$)\\
\hline
0.0 &   0.72 &   0.49  &  ...  & ...\\
0.1 &   0.74 &   0.53  &  ...  & ...\\
0.2 &   0.77 &   0.57  &  ...  & ...\\
0.3 &   0.79 &   0.62  &  ...  & ...\\
0.4 &   0.82 &   0.67  &  ...  & ...\\
0.5 &   0.85 &   0.73  &  ...  & ...\\
0.6 &   0.88 &   0.80  &  ...  & ...\\
0.7 &   0.91 &   0.88  &  ...  & ...\\
0.8 &   0.94 &   0.97  &  ...  & ...\\
0.9 &   0.97 &   ...   &  ...  & ...\\
1.0 &   1.00 &   ...   &  1.00 & ...\\
\hline
\end{tabular}
\end{table}

\subsubsection{Critical mass ratios when the donor is the less massive star}

As far as the donor is the less massive component and provided that
$\dot{M}_{\mathrm{2}}=-\dot{M}_{\mathrm{1}}+\dot{m}$ holds, the following relations are
valid for the mass accretion and the ML rates, $\dot{M}_{\mathrm{1}}$ and $\dot{m}$,
respectively:

\begin{equation}
\label{equation:33}
\dot{M}_{\mathrm{1}}=-\beta\dot{M}_{\mathrm{2}},
\end{equation}
and
\begin{equation}
\label{equation:34} \dot{m}=(1-\beta)\dot{M}_{\mathrm{2}}.
\end{equation}

In the first subcase (hot-spot-driven ML and AML from the primary), we combine
Eqs.~(\ref{equation:33}) and~(\ref{equation:34}) to express $\dot{M}_{\mathrm{2}}$ and
$\dot{m}_{\mathrm{hs}}$ as a function of the mass accretion rate $\dot{M}_{\mathrm{1}}$
and, subsequently, Eq.~(\ref{equation:2}) is written as

\begin{eqnarray}
\label{equation:35}
\frac{\dot{P}}{P}=[-3(q-1)&-&3\sqrt{(q+1)r_{\mathrm{r}}}+\frac{3}{\beta}\cdot(q+1)r_{\mathrm{L}}^{2}\nonumber
\\&-&\frac{1-\beta}{\beta}\cdot\frac{3q^{2}-2q-3}{q+1}]\frac{\dot{M}_{\mathrm{1}}}{M_{\mathrm{2}}}.
\end{eqnarray}

In the second subcase (ML and AML from L3), both the accretion and the ML rate,
$\dot{M}_{\mathrm{2}}$ and $\dot{m}_{\mathrm{L3}}$ respectively, are written as a function
of $\dot{M}_{\mathrm{1}}$, modifying Eq.~(\ref{equation:2}) as

\begin{eqnarray}
\label{equation:36}
\frac{\dot{P}}{P}=[-3(q-1)&-&3\sqrt{(q+1)r_{\mathrm{r}}}+\frac{3}{\beta}\cdot(q+1)r_{\mathrm{L}}^{2}\nonumber
\\&-&\frac{1-\beta}{\beta}\cdot\frac{3j_{\mathrm{L3}}(q+1)^{2}-2q-3}{q+1}]\frac{\dot{M}_{\mathrm{1}}}{M_{\mathrm{2}}}.
\end{eqnarray}

\begin{table}
\caption{Critical mass ratios $q_{\mathrm{cr}}$ as a function of
the degree of liberalism $\beta$ when the donor is the less
massive component (see text for details).} \centering
\label{table:9}
\begin{tabular}{c c c | c c}
\hline \hline \multicolumn{3}{c |}{ML through a hot spot}&
\multicolumn{2}{c}{ML through the L3 point}\\
$\beta$ & $q_{\mathrm{cr}}$ & $q_{\mathrm{cr}}$ & $q_{\mathrm{cr}}$ & $q_{\mathrm{cr}}$\\
$$ & ($r_{\mathrm{r}},r_{\mathrm{L}}=0$) & ($r_{\mathrm{r}},r_{\mathrm{L}}\neq0$) &
($r_{\mathrm{r}},r_{\mathrm{L}}=0$) & ($r_{\mathrm{r}},r_{\mathrm{L}}\neq0$)\\
\hline
0.0 &   ...  &   ...  &  $<0.05$ & $<0.05$ \\
0.1 &   ...  &   ...  &  0.05    & $<0.05$ \\
0.2 &   ...  &   ...  &  0.10    &  0.07   \\
0.3 &   ...  &   ...  &  0.17    &  0.11   \\
0.4 &   ...  &   ...  &  0.24    &  0.17   \\
0.5 &   ...  &   ...  &  0.32    &  0.23   \\
0.6 &   ...  &   ...  &  0.42    &  0.31   \\
0.7 &   ...  &   ...  &  0.53    &  0.40   \\
0.8 &   ...  &   ...  &  0.66    &  0.52   \\
0.9 &   ...  &   0.93 &  0.81    &  0.66   \\
1.0 &   1.00 &   0.83 &  1.00    &  0.83   \\
\hline
\end{tabular}
\end{table}

The numerical expression within the brackets on the r.h.s. of Eqs.~(\ref{equation:35})
and~(\ref{equation:36}) depends exclusively on the mass ratio $q$ of the system. Critical
mass ratio values $q_{\mathrm{cr}}$ then emerge by determining the roots of this expression
for different $\beta$ values, considering a transient disk around the primary, with
$r_{\mathrm{r}}$ and $r_{\mathrm{L}}$ being estimated via Eqs.~(\ref{equation:17})
and~(\ref{equation:19}), respectively. Accordingly, the $q_{\mathrm{cr}}$ values in the
absence of a disk were derived by posing $r_{\mathrm{r}}=r_{\mathrm{L}}=0$. Results for
$q_{\mathrm{cr}}$ values are then given in Table~\ref{table:9}. Systems with mass ratios
$q$ greater than the listed $q_{\mathrm{cr}}$ values are expected to show a decreasing
orbital period (and a concave ETV diagram). The same procedure can be applied by
substituting $r_{\mathrm{1}}$ for $r_{\mathrm{r}}$ to find the corresponding
$q_{\mathrm{cr}}$ values in the presence of a permanent disk.

\begin{figure}
\resizebox{\hsize}{!}{\includegraphics{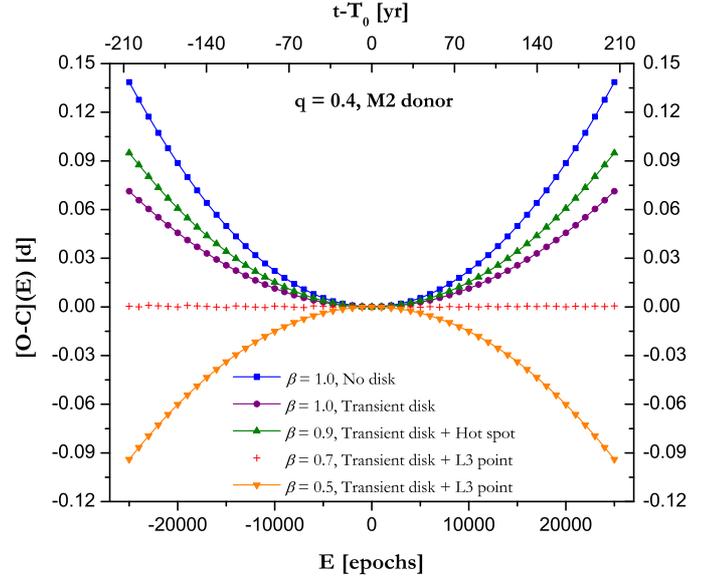}}
\caption{Synthetic ETV diagrams for a $q = 0.4$ semi-detached
system when the donor is the secondary star (step $\equiv$ 1000
cycles). Squares and circles account for a conservative era
($\beta = 1.0$) either in the absence or in the presence of a
transient disk, respectively. Upwardly pointing triangles
($\beta = 0.9$) account for a hot-spot-driven non-conservative
era in the presence of a transient disk, while crosses
($\beta = 0.7$) and downwardly pointing triangles
($\beta = 0.5$) represent a non-conservative era where material
escapes via the L3 point. (This figure is available in color in
the electronic form.)
              }
         \label{figure:3}
   \end{figure}

In Fig.~\ref{figure:3}, synthetic ETV diagrams have been constructed by means of
Eqs.~(\ref{equation:15}), (\ref{equation:35}), and (\ref{equation:36}) for a $q = 0.4$
semi-detached system with a primary gainer of $r_{\mathrm{1}}=0.13$ when the mass accretion
rate is fixed at $\dot{M}_{\mathrm{1}}=10^{-8}\,M_{\mathrm{\odot}}\,\mathrm{yr}^{-1}$. The
geometry of the anticipated ETV diagram for the same system is modulated according to the MT
mode and the degree of the liberalism (i.e., the $\beta$ value). Indeed, for the system
\object{RR Dra}, which is suspected of accommodating a transient disk orbiting the primary star
(Kaitchuck et al. \cite{KHS85}), a conservative MT rate close to
$6\times10^{-7}\,M_{\mathrm{\odot}}\,\mathrm{yr}^{-1}$ is derived, while a value of $3\times10^{-7}\,M_{\mathrm{\odot}}\,\mathrm{yr}^{-1}$ is (under)estimated by the
conventional approach (Zasche et al. \cite{ZLWN08}). Moreover, the observed period variations
of \object{X Tri} could be ascribed to a non-conservative MT process with ML rates through L3 ranging
from $-2\times10^{-8}$ to $-6\times10^{-8}\,M_{\mathrm{\odot}}\,\mathrm{yr}^{-1}$ (see also
Nanouris et al. \cite{NKARL13}), whereas the conservative approach fails to provide any
reasonable explanation (Rovithis-Livaniou et al. \cite{RLKRA00}).

\subsection{Case IV: Angular momentum loss through gravitational radiation}

The last studied individual case concerns the effects of gravitational wave radiation (GWR), a
physical mechanism that is highly suspected to drive binaries evolution with mainly degenerate
components (e.g., Kramer et al. \cite{KSM06}; Antoniadis et al. \cite{AFW13}). In non-degenerate
binaries, GWR seems to become a significant source of AML once binaries reach the contact phase
(Webbink \cite{W76}).

The rate at which a binary loses angular momentum because of the GWR mechanism is given by the
well-known relation of Landau \& Lifshitz (\cite{LL62}):

\begin{equation}
\label{equation:37} \dot{J}_{\mathrm{g}}=-\frac{32\left(2\pi
G\right)^{7/3}}{5c^{5}} \cdot
\frac{(M_{\mathrm{1}}M_{\mathrm{2}})^{2}}{(M_{\mathrm{1}}+M_{\mathrm{2}})^{2/3}P^{7/3}},
\end{equation}
where $c$ is the speed of light in a vacuum. By ignoring MT and ML terms in
Eqs.~(\ref{equation:1}) and~(\ref{equation:2}) and disregarding any other possible AML processes,
the following differential equation is set and is solved for the orbital period $P(t)$:

\begin{equation}
\label{equation:38} \dot{P}=-b_{\mathrm{g}}P^{-\frac{5}{3}},
\end{equation}
where the constant coefficient $b_{\mathrm{g}}$ is given by the relation

\begin{equation}
\label{equation:39} b_{\mathrm{g}}=
\frac{96(2\pi)^{8/3}G^{5/3}}{5c^{5}}\cdot\frac{M_{\mathrm{1}}M_{\mathrm{2}}}{(M_{\mathrm{1}}+M_{\mathrm{2}})^{1/3}}>0.
\end{equation}

The solution of Eq.~(\ref{equation:38}) under the initial condition
$(t,P(t))=(T_\mathrm{0},P_\mathrm{e})$ is

\begin{equation}
\label{equation:40}
P(t)=\left[P_{\mathrm{e}}^{\frac{8}{3}}-\frac{8}{3}b_{\mathrm{g}}(t-T_{\mathrm{0}})\right]^{\frac{3}{8}}.
\end{equation}
Then, integration of Eq.~(\ref{equation:4}) under the initial condition
$(\epsilon,t)=(0,T_\mathrm{0})$ yields

\begin{equation}
\label{equation:41}
t-T_\mathrm{0}=\frac{3}{8b_{\mathrm{g}}}\left[P_{\mathrm{e}}^{\frac{8}{3}}-\left(P_{\mathrm{e}}^{\frac{5}{3}}
-\frac{5}{3}b_{\mathrm{g}}\epsilon\right)^{\frac{8}{5}}\right].
\end{equation}
Hence, through Eqs.~(\ref{equation:40}) and~(\ref{equation:41}), the orbital period $P(\epsilon)$
as a function of the continuous time variable $\epsilon$ is

\begin{equation}
\label{equation:42}
P(\epsilon)=\left(P_{\mathrm{e}}^{\frac{5}{3}}-\frac{5}{3}b_{\mathrm{g}}\epsilon\right)^{\frac{3}{5}}.
\end{equation}

Integration of Eq.~(\ref{equation:5}) under the initial condition
$(\epsilon,\Delta T(\epsilon))=(0,0)$ eventually gives the $\Delta T(\epsilon)$ function

\begin{equation}
\label{equation:43} \Delta
T(\epsilon)=\frac{3}{8b_{\mathrm{g}}}\left[P_{\mathrm{e}}^{\frac{8}{3}}-\left(P_{\mathrm{e}}^{\frac{5}{3}}
-\frac{5}{3}b_{\mathrm{g}}\epsilon\right)^{\frac{8}{5}}\right]-P_{\mathrm{e}}\epsilon,
\end{equation}
which practically pictures the theoretically anticipated form of the ETV diagram under the
action of GWR-AML, and is proved to be concave for any cycle $\epsilon$, a result referring
to the continuously decreasing orbital period function $P(\epsilon)$, expressed by
Eq.~(\ref{equation:42}).

\begin{table}
\caption{Minimum time spans at $\varepsilon = 0.001\,\mathrm{d}$
and 0.01\,d for several systems that lose angular momentum
through gravitational radiation.} \centering \label{table:10}
\begin{tabular}{l c c c c c}
\hline \hline \multicolumn{5}{c}{$t-T_{\mathrm{0}}$}\\
\multicolumn{5}{c}{$[yr]$}\\
\hline
System & $P_{\mathrm{e}}$ & $M_{\mathrm{1}}+M_{\mathrm{2}}$ & \multicolumn{2}{c}{O--C noise level}\\
& $[d]$ & $[M_{\mathrm{\odot}}]$ & $\varepsilon = 0.001\,\mathrm{d}$ & $\varepsilon = 0.01\,\mathrm{d}$\\
\hline
...    &  1.00 &   1+1           &  $933$ &  $2951$ \\
...    &  1.00 &   5+5           &  $244$ &  $772 $ \\
...    &  1.00 &   8+8           &  $165$ &  $521 $ \\
CC Com &  0.22 &   0.7+0.4$^{a}$ &  $212$ &  $670 $ \\
W UMa  &  0.33 &   1.2+0.6$^{a}$ &  $246$ &  $779 $ \\
AH Cep &  1.77 &   16+13$^{b}$   &  $216$ &  $684 $ \\
WR 20a &  3.69 &   83+82$^{c}$   &  $135$ &  $426 $ \\
\hline
\multicolumn{5}{l}{$^a$Hilditch et al. (\cite{HKM88})}\\
\multicolumn{5}{l}{$^b$Burkholder et al. (\cite{BMM97})}\\
\multicolumn{5}{l}{$^c$Bonanos et al. (\cite{BSU04})}
\end{tabular}
\end{table}

The traceability of a $1+1\,M_{\mathrm{\odot}}$, a $5+5\,M_{\mathrm{\odot}}$, and a
$8+8\,M_{\mathrm{\odot}}$ system are then examined, all with $P_{\mathrm{e}}$ = 1\,d. As
seen in Table~\ref{table:10}, even for a quite massive pair of $8+8\,M_{\mathrm{\odot}}$,
it is almost impossible to observe any GWR-driven O--C variations, since at least 160 and
500\,yr of dense monitoring are needed below noise levels equal to 0.001 and 0.01\,d,
respectively. In Table~\ref{table:10}, we also tabulate the corresponding time intervals $t(\epsilon_{\mathrm{min}})-T_{\mathrm{0}}$ expected for the short-period contact binaries
\object{CC Com} and \object{W UMa} ($P_{\mathrm{e}} <$ 0.4\,d) and, beyond these, for the
two massive binaries \object{AH Cep} and \object{WR 20a}
($M_{\mathrm{1}}+M_{\mathrm{2}} > 30\,M_{\mathrm{\odot}}$). Even for \object{WR 20a}, which
is among the most massive binaries known to date (Bonanos et al. \cite{BSU04}), the GWR
effects on its ETV diagram would only become detectable after 135\,yr of accurate recorded
times of minima.

\begin{figure}
\resizebox{\hsize}{!}{\includegraphics{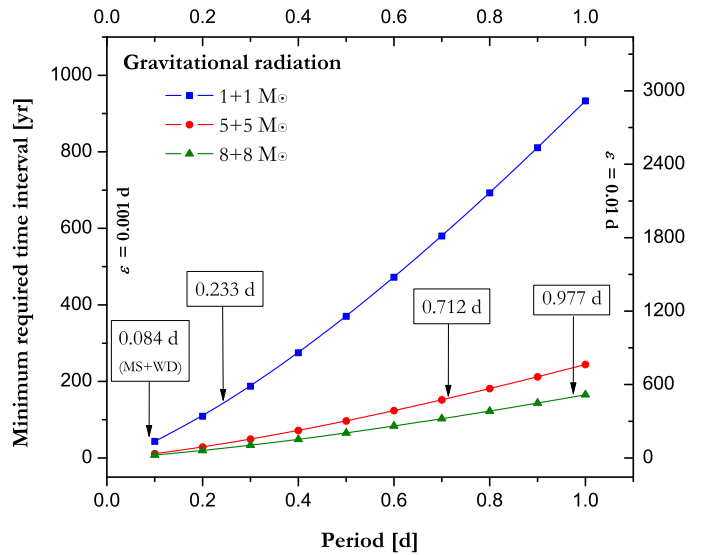}}
\caption{Minimum time spans considering that gravitational
radiation is the driving evolutionary process of a
$1+1\,M_{\mathrm{\odot}}$ (squares), a $5+5\,M_{\mathrm{\odot}}$
(circles), and a $8+8\,M_{\mathrm{\odot}}$ system (up triangles)
with various periods for noise levels equal to 0.001\,d (left
vertical axis) and 0.01\,d (right vertical axis). The labels
depict the period thresholds at which each system attains the
contact configuration. (This figure is available in color in the
electronic form.)
              }
         \label{figure:4}
   \end{figure}

An additional investigation is also performed to find out whether GWR-driven modulations are
visible in a wide range of periods, varying from 0.1 to 1.0\,d. All three hypothetical systems
are explored, keeping in mind that a period threshold should exist for each case as a response
to the contact configuration (Fig.~\ref{figure:4}). The critical limit of the orbital period
$P_{\mathrm{cr}}$ was determined by considering radii of $3.6\,R_{\mathrm{\odot}}$ and
$5.2\,R_{\mathrm{\odot}}$ for a $5\,M_{\mathrm{\odot}}$ and a $8\,M_{\mathrm{\odot}}$ dwarf
star, respectively. Especially in the case of the $1+1\,M_{\mathrm{\odot}}$ system, the critical
period $P_{\mathrm{cr}}$ was also determined considering that one member of the pair is a white
dwarf with radius equal to $0.01\,R_{\mathrm{\odot}}$ (MS+WD).

A close inspection of Fig.~\ref{figure:4} reveals that, as in the magnetic braking case (see
NKARL11), the shorter the orbital period, the shorter the time required. Close to the period
threshold of each examined system, slightly less than two centuries of observations are needed
for producing measurable O--C variations. However, the MS+WD system seems to be an exceptional
case that is likely to be highly detectable, demanding only five decades of monitoring at most.
This finding mainly accommodates the group of eclipsing cataclysmic binaries with orbital periods
of a few hours (see, e.g., Pilar\v{c}\'{i}k et al. \cite{PWDHK12} and references therein).

\section{Discussion}

We examined the orbital evolution and the associated synthetic ETV diagrams of binary systems
that experience some of the most common manifestations of MT mechanisms in short, human life
timescales. A complementary study was also carried out for GWR-induced modulations. The minimum
range, required for the detectability of the ensuing O--C variations, was then estimated following
the methodology introduced in Nanouris et al. (\cite{NKARL11}) for a variety of both physical and
observational parameters where the procedure had been proved to be sensitive.

The conservative MT process was thoroughly investigated, and it was shown that for the direct-impact
mode in the presence of a transient disk -- which extracts large amounts of the orbital angular
momentum -- around the more massive component there is a critical mass ratio
($q_{\mathrm{cr}} \approx 0.83$) above which the period turns out to decrease, contrary to those
expectations that emerged by the conventional approach.

Different non-conservative MT modes were also examined by employing a parameter $\beta$ to denote
the degree of conservatism. Values of $\beta$ close to zero account for violent non-conservative
eras, while values close to unity account for highly conservative ones. Taking $\beta$ as a free
parameter in their hot-spot-driven evolutionary models, van Rensbergen et al. (\cite{VRDGDLM08})
have shown that short non-conservative eras soon after an RLOF ignition may take place with a
variable $\beta$, even reaching the value of $\beta \sim 0.2$. The present study revealed that in
a non-conservative MT process induced by a hot-spot emission, AML is too weak to drive the orbital
evolution, still allowing an increasing period even in strong but reasonable non-conservative eras
(i.e., $\beta \leq 0.4$), however only for a narrow range of high mass ratios (i.e., $q \geq 0.7$)
in systems where the donor is the primary star.

Instead, ML via the outer points L2 or L3 seems to be a strong AML process, capable of reducing
the period at a very rapid pace. Sytov et al. (\cite{SKBKB07}) have shown that the efficiency of
accumulation in accretion disks of close binaries is not higher than 20-30\% for typical values of
turbulent viscosity, while it hardly reaches a rate of 50\%, provided that the magnetic field is
taken into consideration (Bisikalo \cite{B10}), impelling ML through L3 at significant amounts. We
showed that the period is expected to decrease even during weak non-conservative eras through the
L2 point, while when L3 is the libration point, it was revealed that there is always a critical mass
ratio $q_{\mathrm{cr}}$ above which the period is expected to decrease. The critical values
$q_{\mathrm{cr}}$ decrease as the degree of liberalism is progressively widened. Unlike the hot-spot
emission, this allows a period reduction even for very low mass ratios (i.e., $q \leq 0.3$) in
systems where the donor is the secondary star.

The above findings prove that ML through the L3 point is a forceful candidate mechanism for
interpreting the concave ETV diagrams that represent semi-detached systems with the less massive star
to be the contact component (see, e.g., Yang \& Wei \cite{YW09} for an updated catalog of binaries
that display this peculiar behavior). Among them, \object{AF Gem}, \object{TU Her}, \object{AT Peg},
\object{TY Peg}, \object{Z Per}, \object{Y Psc}, and \object{X Tri} are systems with low mass ratios,
and consequently, they are strongly suspected to undergo a non-conservative MT era, accompanied by ML
through the L3 point at a significant rate. Recently, light curves of several early-type systems with
an accretion disk have been analyzed using the new improved code of
Djura\v{s}evi\'{c} et al. (\cite{DRLRBB05}; \cite{DVA08}) for estimating the physical disk parameters.
Their analysis has indicated the presence of a bright spot, a region on the disk edge opposite (and
close) to the L3 point, through which matter is expected to escape from the system. \object{DL Cyg},
\object{$\beta$ Lyr}, \object{AU Mon}, \object{V393 Sco}, \object{RY Sct}, and \object{DQ Vel} are low
mass ratio systems that could experience a non-conservative MT era.

It is noteworthy that the MT process is capable of driving the orbital evolution in the limited temporal
range of ETV diagrams at rates similar to the wind-driven ML counterparts (NKARL11), implying that these
two mechanisms might be strongly competitive (see also Erdem et al. \cite{EBD05}; Erdem et al.
\cite{EDBD07a},b). Consequently, ML through stellar winds should not be neglected in systems of
semi-detached or contact configuration. Besides, the inferred critical mass ratios from the present
analysis are expected to change considerably when this kind of ML is involved in our calculations.
Evidently, the simplistic $\beta-q$ schemes seem to be insufficient in describing the short orbital
evolution of a binary in which MT is not the leading evolutionary mechanism. This may be consistent with
the findings of Qian (\cite{Q01}), who has shown that a critical value close to
$q_{\mathrm{cr}} \approx 0.40$ appears to invert the monotonicity of the observed orbital evolution in
contact binaries. However, this did not emerge clearly from the present study. The determination of the
period monotonicity in the presence of numerous physical processes of a different nature, although
feasible, is rather complicated and requires a more careful investigation.

The sufficiency of the commonly adopted parabolic representation of the long-term variations in ETV
diagrams will be examined in a forthcoming paper through Taylorian expansions of the analytic
$\Delta T(\epsilon)$ expressions.

\begin{acknowledgements}

      This project was funded by the State Scholarships Foundation of Greece
      (IKY) as part of the first author's Ph.D. thesis and financially supported
      by the Special Account for Research Grants (ELKE), No. 70/4/9702, of the
      National \& Kapodistrian University of Athens, Greece. N. N. gratefully
      acknowledges them for their support. N. N. also acknowledges financial
      support under the ``MAWFC" project. Project MAWFC is implemented under the
      ``ARISTEIA II" action of the ``OPERATIONAL PROGRAMME EDUCATION AND LIFELONG
      LEARNING". The project is cofunded by the European Social Fund (ESF) and
      National Resources. We are thankful to the referee whose precious
      recommendations contributed to the improvement of both the methodology and
      the clarity of the manuscript. We would also like to thank Dr. A. Chiotellis
      for many valuable discussions and suggestions dealing with the orbital
      evolution of binary systems. This research has made use of SIMBAD database,
      operated at the CDS (Strasbourg, France) and NASA's Astrophysics Data System
      Bibliographic Services.

\end{acknowledgements}

\end{document}